\documentclass[a4paper,11pt]{article}
\usepackage{jheppub} 
\usepackage{lineno}
\usepackage{subfigure}
\usepackage{graphicx}
\usepackage{subcaption}
\usepackage{caption}
\usepackage{cancel}
\usepackage{comment}

\usepackage{xcolor}
\usepackage[normalem]{ulem}

\newcommand{\ARdel}[1]{{\color{red}[X]}}

\newcommand{\KGdel}[1]{{}}

\newcommand{\BEdel}[1]{{\color{teal}[X]}}

\preprint{CERN-TH-2025-211}
\arxivnumber{2510.21922} 

\title{\boldmath Constrained instantons in scalar field theories}

\author[a]{Benjamin Elder,}
\author[a]{Kinga Gawrych,}
\author[a,b]{and Arttu Rajantie.}
\affiliation[a]{Abdus Salam Centre for Theoretical Physics, Imperial College London, London SW7 2AZ, United Kingdom}
\affiliation[b]{Theoretical Physics Department, CERN, 1211 Geneva 23, Switzerland}

\emailAdd{k.gawrych22@imperial.ac.uk}

\abstract{
Instantons, localised saddle points of the action, play an important role in describing non-perturbative aspects of quantum field theories, for example vacuum decay or violation of conservation laws associated with anomalous symmetries. However, there are theories in which no saddle point exists. In this paper, we revisit the idea of constrained instantons, proposed initially by Affleck in 1981, and develop it into a complete method for computing the vacuum decay rate in such cases. We apply this approach to the massive scalar field theory with a negative quartic self-interaction using two different constraints. We solve the field equations numerically and find a two-branch structure, with two distinct solutions for each value of the constraint. By counting the negative modes, we identify one branch of solutions as the constrained instantons and the other as the minima of the action subject to the constraint. We discuss their significance for the computation of the vacuum decay rate.}

\begin{document}
\maketitle
\flushbottom

\section{Introduction}\label{sec:intro}

Many quantum field theories predict phenomena that cannot be captured by perturbation theory, for example baryon number violating processes in the Standard Model, or vacuum transitions between vacuum states. These can often be described by instantons, classical solutions of the field equations localised around a point in the Euclidean spacetime, which correspond to saddle points of the action and
describe quantum tunneling between different vacuum states.
Therefore these solutions play a crucial role in our understanding of quantum field theory.

A classic example is vacuum decay. It
occurs typically in theories where the potential has a local minimum, termed
the \emph{false vacuum}. In classical field theory a local minimum is stable, but in quantum field theory the field is allowed to tunnel from the false vacuum to the \emph{true vacuum}, i.e., the global minimum. When a corresponding instanton solution exists, its action determines the rate of this transition~\cite{Coleman:1977py, Callan:1977pt}.

However, sometimes theories with metastable vacua do not, in fact, have any instanton solutions~\cite{Affleck:1980mp}. One important example of this phenomenon concerns the Yang-Mills instanton~\cite{Belavin:1975fg}, which describes tunneling between vacua with different Chern-Simons number. In the pure Yang-Mills theory, the solution can be found and its action calculated analytically.
Because of the scale invariance of the classical theory, there is, in fact, a one-parameter family of these solutions, all with the same action, which are related by a scaling transformation.
Similar transitions between different Chern-Simons numbers also exist in the electroweak theory, where the chiral anomaly~\cite{Adler:1969gk,Bell:1969ts} implies that they violate baryon number conservation \cite{tHooft:1976rip}. However, because of the presence of the Higgs field, a simple scaling argument shows that there can be no non-trivial stationary points of the action in the electroweak theory, and therefore no instanton solutions~\cite{Derrick:1964ww}.

A simpler example, and the topic of this work, is the real $\phi^4$ scalar field theory with a negative self-coupling. With a positive mass term, the theory has a local minimum at $\phi=0$, and even though the potential is not bounded from below and has therefore no global minimum, one can still ask what the lifetime of this false vacuum is. In fact, in the massless limit, in which there is no local minimum, one can find the relevant instanton solution analytically~\cite{Fubini:1976jm,Lee:1985uv}. 
Because it has a finite action, it implies that even though $\phi=0$ is classically unstable, it is metastable in quantum field theory.
As in the case of the Yang-Mills theory, scale invariance again implies that there is a one-parameter family of such solutions in the massless theory, and the introduction of a mass term breaks their degeneracy so that no saddle point solution can exist~\cite{Affleck:1980mp}.

The question we address in this paper is how one can compute the vacuum decay rate in such a case, where no instanton solution exists.
Often, the lack of instanton solutions can be traced back to the existence of a continuous transformation that reduces the action of any field configuration monotonically. In the two specific cases discussed above, the transformations are spatial rescalings of the solution. In the massless theory, which has the one-parameter family of instanton solutions, these same scaling transformations map the solutions to each other, which means that in that limit, it is possible to parameterise them by their spatial size.

This topic was originally studied for scalar $\phi^4$ and Yang-Mills-Higgs theories by Affleck in~\cite{Affleck:1980mp}, inspired by earlier related work by 't~Hooft~\cite{tHooft:1976snw} and Frishman and Yankielowicz \cite{PhysRevD.19.540}. 
His idea was to fix the instanton size by introducing a constraint into the path integral.
This singles out one of the massless instantons as the zeroth-order solution, to which he then computed corrections perturbatively.
He referred to these configurations as \emph{constrained instantons}.

The constrained instanton method was used in Yang-Mills-Higgs theory to compute several baryon- and lepton-number-violating cross-sections in high energy processes \cite{Silvestrov:1992ct, Espinosa:1989qn, osti_6099255, Ringwald:1989ee, Khoze:1990bm}. This approach was revisited multiple times in the 1990s, in particular by Klinkhamer \cite{Klinkhamer:1991pq, Klinkhamer:1993kn, Klinkhamer:1994xw, Klinkhamer:1996ad, Klinkhamer:1997pd}, and improved upon by Gibbs \cite{Gibbs:1995xt} by including the effect of the $U(1)$ field on the instanton solutions. There has also been work on explicit construction of constrained instantons in scalar field theory \cite{PhysRevD.61.105020}. Alternative methods for studying vacuum decay in theories with no instantons, such as the valley instanton approach \cite{Aoyama:1995ca, Aoyama:1995zt, Wang:1994rz} and the tunneling potential approach \cite{Espinosa:2018hue,Espinosa:2019hbm,Espinosa:2022jlx,Espinosa:2024ufg,Espinosa:2025ejf} have also been 
proposed. Recently, constrained instantons have been used in the context of gravity \cite{PhysRevD.104.L081501}, and as a tool for constructing Skyrmion-like solutions in theories where usual Skyrmions are forbidden \cite{GarciaMartin-Caro:2022ukd}. 

Despite these successes, the constrained instanton approach has so far been limited to cases in which the constrained instantons are small perturbations of the instanton solutions of the massless theory\footnote{The authors are aware of one attempt at a non-perturbative construction of a constrained instanton in $SU(2)$ Yang-Mills-Higgs theory --- see refs. \cite{Klinkhamer:1993kn,Klinkhamer:1996ad}.}. This is manifested in the use of the instanton ``size'' -- a concept restricted to the massless theory -- to determine the relevant solution, and in the use of perturbation theory to find the solution. 
In this paper we present a fully non-perturbative formulation of the idea, which allows us to find the relevant solutions even when they are not well approximated by massless instantons. We also obtain an explicit mathematical expression for the vacuum decay rate in terms of these constrained instanton solutions in a generic scalar field theory. We apply our method to a simple toy model, in which we obtain the constrained instanton solutions.

We begin section~\ref{sec:constinst} with a review of the physics of vacuum decay in theories with exact instanton solutions. We then show how constrained instantons can be used to compute the decay rate in theories with no instantons, and how the use of Lagrange multipliers helps simplify the calculation. We also derive an analytic expression for the constrained instanton rate. In section~\ref{sec:massphi} we apply our improved approach to the case of a massive scalar field with negative quartic term in the potential and we discuss the allowed forms of the constraint. We also describe a method of counting the number of negative modes in the spectrum of a given solution in the presence of a constraint, which is crucial to determine whether it is in fact a constrained instanton, and therefore whether it contributes to the decay rate. We find the constrained solutions and their action numerically in section~\ref{sec:results} for two different types of constraint, allowing us to highlight the similarities and differences due to two reasonable but distinct choices of constraint, but we leave the numerical computation of the full decay rate for future work. Finally, we conclude in section~\ref{sec:concl}.

\section{Constrained instantons}\label{sec:constinst}

\subsection{Vacuum decay rate}\label{sec:vacuumdecay}

Let us consider a theory with some fields, denoted by $\phi$, and action $S[\phi]$. We assume that the action has a local minimum, in which the fields have a constant value $\phi^0$. We refer to this local minimum as the {\it false vacuum}. In many cases, the action also has a global minimum which would be called the true vacuum, but at least formally it is possible to also consider theories in which no such global minimum exists.

In classical field theory, the false vacuum is stable against small perturbations, but in quantum field theory it is only metastable, as the field can escape the local minimum through quantum tunnelling. The metastability of the false vacuum is characterised by the vacuum decay rate $\Gamma$ defined as the decay probability per spacetime volume.

Denoting the false vacuum quantum state by $|0\rangle$,
the vacuum decay rate is related to the vacuum persistence probability,
\begin{equation}
    \label{equ:vacpersist}
    \left|\left\langle 0\right|e^{i\hat{H}t}\left|0\right\rangle\right|^2
    \sim e^{-\Gamma Vt}~,
\end{equation}
where $\hat{H}$ is the Hamiltonian and $V$ is the volume of space.
Inverting this gives
\begin{equation}
\label{equ:GammaPI}
    \Gamma=
    -\frac{2}{V}\lim_{t\rightarrow\infty}
    \frac{1}{t}{\rm Re}\log \left\langle 0\right|e^{i\hat{H}t}\left|0\right\rangle
    =
    \frac{2}{V}\lim_{\tau\rightarrow\infty}
    \frac{1}{\tau}{\rm Im}\log \left\langle 0\right|e^{-\hat{H}\tau}\left|0\right\rangle
    =
    \frac{2}{\cal V}{\rm Im}\log 
    \int {\cal D}\phi e^{-S_E[\phi]}~,
\end{equation}
where we carried out the Wick rotation $t=i\tau$ and in the final form, 
${\cal V}=V\tau$ is the spacetime volume. $S_E$ is the Euclidean action and the boundary conditions of the Euclidean path integral correspond to $\phi=\phi^0$ at infinity.

In many cases, when the theory is weakly coupled, one can use the saddle point approximation to evaluate the path integral \eqref{equ:GammaPI},
\begin{equation}
\label{equ:saddle}
    \int {\cal D}\phi e^{-S_E[\phi]}
    \approx 
Z[\phi^0]+\sum_{\hat\phi}Z[\hat\phi]~,
    \quad
Z[\phi]=
    \left(\text{Det}\, M[\phi]
    \right)^{-1/2}e^{-S_E[\phi]}~,
\end{equation}
where $\hat\phi$ labels the stationary points of the action which satisfy the correct boundary conditions, and $\text{Det}\, M[\phi]$ is the functional determinant of the fluctuation operator
\begin{equation}
\label{equ:fluctop}
    M(x,y)=\frac{\delta^2 S}{\delta \phi(x)\delta \phi(y)}~,
\end{equation}
calculated at $\phi$.

In order to compute eq.~\eqref{equ:GammaPI}, we are interested in stationary points that would give an imaginary contribution. This means that the fluctuation determinant has to be negative, and therefore the fluctuation operator \eqref{equ:fluctop} has to have an odd number of negative modes. In practice, the dominant contribution usually comes from \emph{instantons} --- saddle point configurations with a single negative mode which are localised in spacetime.\footnote{Topological instantons such as the Yang-Mills instanton are an exception, as they do not have negative modes.} Such a configuration breaks translation invariance, which means that it has one zero mode per spacetime dimension. Integration over these zero modes using collective coordinates gives a factor proportional to the spacetime volume, so by defining $\overline{\text{Det}}$ as the functional determinant with these zero modes removed, we have \cite{Callan:1977pt}
\begin{equation}
\label{equ:saddle2}
    \int {\cal D}\phi e^{-S_E[\phi]}
    \approx 
    Z[\phi^0]
    \left[
        1 + 
        i\frac{{\cal V}}{2}
        \,\left(\frac{S[\phi^{\rm inst}]}{2\pi}\right)^2
        \left|
        \frac{{\overline{\text{Det}}}\,M[\phi^{\rm inst}]}
        {\text{Det}\,M[\phi^0]}
        \right|^{-1/2}e^{-\left(S_E[\phi^{\rm inst}]-S_E[\phi^0]\right)}
    \right]~,
\end{equation}
Substituting this into eq.~\eqref{equ:GammaPI} gives
\begin{equation}
\label{equ:instGamma}
\Gamma\approx
    \KGdel{2}\left(\frac{S[\phi^{\rm inst}]}{2\pi}\right)^2\left|
        \frac{{\overline{\text{Det}}}\, M[\phi^{\rm inst}]}
        {\text{Det}\, M[\phi^0]}
        \right|^{-1/2}e^{-\left(S_E[\phi^{\rm inst}]-S_E[\phi^0]\right)}~.
\end{equation}

There are, however, some cases when an instanton solution does not exist. This happens when there is a continuous transformation $\phi\rightarrow \phi'$ which lowers the action of any non-trivial configuration, i.e., $S_E[\phi']<S_E[\phi]$ for any $\phi\ne\phi^0$.
In that case, the action cannot have any stationary points other than $\phi^0$. For example, in the Yang-Mills Higgs theory
and the scalar field theory discussed in section~\ref{sec:massphi},
this is achieved by a scaling transformation.
To generalise the instanton approach to these cases, Affleck~\cite{Affleck:1980mp} introduced the concept of a {\it constrained instanton}.

Affleck defined his constrained instantons in terms of the instanton ``size'', but we generalise this by defining a constraint functional $\xi[\phi]$ which is chosen in such a way that it is not invariant under the scaling transformation. For concreteness, we assume that it has the form of an integral over some local function ${\cal O}$ of the field and its derivatives,
\begin{equation}\label{eq:constraint}
    \xi[\phi]=\int d^4x\, {\cal O}~.
\end{equation}

We can now insert a delta functional in the path integral and integrate over it,
\begin{equation}
\label{equ:constPI}
    \int {\cal D}\phi e^{-S_E[\phi]}
    =\int d\bar\xi \int {\cal D}\phi \,\delta\left(\xi[\phi]-\bar\xi\right) e^{-S_E[\phi]}~.
\end{equation}
Now the path integral is taken only over the space of functions $\phi(x)$ that satisfy the constraint $\xi[\phi]=\bar\xi$, which we denote by $\mathcal{F}_{\bar\xi}$. Even if the action has no saddle point in the space of all functions, it may have one in $\mathcal{F}_{\bar\xi}$, and in that case we can use the saddle point approximation to evaluate the path integral (\ref{equ:constPI}).

Denoting the minimum of the action subject to the constraint by $\phi^0_{\bar\xi}$ and the saddle point by $\phi^{\rm inst}_{\bar\xi},$ we have in analogy with eq.~\eqref{equ:saddle},
\begin{equation}
    \int {\cal D}\phi \,\delta\left(\xi[\phi]-\bar\xi\right) e^{-S_E[\phi]}
    \approx
    Z_{\bar\xi}[\phi^0_{\bar\xi}]
    +Z_{\bar\xi}[\phi^{\rm inst}_{\bar\xi}]~,
    \quad
    Z_{\bar\xi}[\phi]=
    \left({\text{Det}}_{\bar\xi} M_{\bar\xi}[\phi]
    \right)^{-1/2}e^{-S_E[\phi]}~,
\end{equation}
where {$M_{\bar\xi}$ and}
${\text{Det}}_{\bar\xi}$ denote
the fluctuation operator and
the functional determinant in {$\mathcal{F}_{\bar\xi}$}, respectively, and are different from their counterparts $M$ and $\text{Det}$ in the space of all functions.

The full path integral eq.~\eqref{equ:constPI} is therefore approximated as
\begin{equation}\label{eq:fullconPI}
        \int {\cal D}\phi  e^{-S_E[\phi]}
    \approx
    \int d\bar\xi 
    Z_{\bar\xi}[\phi^0_{\bar\xi}]
    +
    \int d\bar\xi 
    Z_{\bar\xi}[\phi^{\rm inst}_{\bar\xi}]~,
\end{equation}
where the first term is real and the second term is imaginary. However, we can note that the real part of the full path integral can be computed directly without using  the constraint, so we can equally well write
\begin{equation}
        \int {\cal D}\phi  e^{-S_E[\phi]}
    \approx
    Z[\phi^0]
    +
    \int d\bar\xi 
    Z_{\bar\xi}[\phi^{\rm inst}_{\bar\xi}]~.
\end{equation}
In analogy with eq.~\eqref{equ:instGamma} we therefore obtain the expression for the vacuum decay rate in terms of the constrained instantons $\phi^{\rm inst}_{\bar\xi}$ as
\begin{equation}
    \label{equ:constGamma}
    \Gamma\approx
    \KGdel{2}
    \int d\bar\xi\,
    \left(\frac{S[\phi^{\rm inst}_{\bar\xi}]}{2\pi}\right)^2
    \left|
        \frac{{\overline{\text{Det}}}_{\bar\xi} M_{\bar\xi}[\phi^{\rm inst}_{\bar\xi}]}
        {{\text{Det}} M[\phi^0]}
        \right|^{-1/2}e^{-\left(S_E[\phi^{\rm inst}_{\bar\xi}]-S_E[\phi^0]\right)}~.
\end{equation}

In order to compute the vacuum decay rate $\Gamma$, one needs to choose the constraint functional $\xi[\phi]$ in such a way that a saddle point solution exists in {$\mathcal{F}_{\bar\xi}$}
for the relevant range of the constraint parameter $\bar\xi$. One then finds these constrained instanton solutions $\phi^{\rm inst}_{\bar\xi}$, calculates the action and the functional determinant, and integrates over $\bar\xi$. In this paper, we will focus on the first two steps, finding the constrained instantons and calculating their action.

From now on, to simplify the notation, we will denote the Euclidean action $S_E[\phi]$ by $S[\phi]$, and the constrained instantons $\phi^{\rm inst}_{\bar\xi}$ by $\phi_{\bar\xi}$.

\subsection{Lagrange multiplier}\label{sec:Lagmult}

In practice, we can find the constrained instantons using the method of Lagrange multipliers. 
We define the \emph{modified action}
\begin{equation}\label{eq:deformact}
    \tilde S_\kappa[\phi] = S[\phi] + \kappa \xi[\phi] = S[\phi] + \kappa \int d^4x \, \mathcal{O}\,,
\end{equation}
where $\kappa$ is a constant Lagrange multiplier. 
Any stationary point $\tilde\phi_\kappa$ of this modified action is then also a stationary point of the original action $S[\phi]$ in the subspace of functions that satisfy the constraint $\xi[\phi]=\xi[\tilde\phi_\kappa]$. This means that the constrained instanton solution $\phi_{\bar\xi}$ for the constraint value $\bar\xi$ is given by 
\begin{equation}
    \label{equ:CIsol}
    \phi_{\bar\xi}=\tilde\phi_\kappa~, \quad \bar\xi=\xi[\tilde\phi_\kappa]~.
\end{equation}

In textbook examples of Lagrange multipliers, one is usually looking for the solution for a specific value $\bar\xi$ of the constraint parameter, and therefore one inverts the second equation of eq.~\eqref{equ:CIsol} to find the value of $\kappa$ it corresponds to. However, because we are interested in all values of $\bar\xi$, we can omit that step. Instead, we use the two equations in eq.~\eqref{equ:CIsol} to determine both $\bar\xi$ and $\phi_{\bar\xi}$ for each value of $\kappa.$

For future reference, it is useful to note some important properties of Lagrange multipliers. First, we have the conjugate relations
\begin{equation}\label{eq:legendrexi}
    \bar\xi=\frac{d \tilde S_\kappa[\tilde\phi_\kappa]}{d \kappa}~,
\end{equation}
    and
\begin{equation}\label{eq:legendrekap}
    \kappa=-\frac{d S[\phi_{\bar\xi}]}{d\bar\xi}~.
\end{equation}
These relations give us a powerful check of the validity of our numerical results. This is discussed in more detail in appendix~\ref{app:CISols}.

Second, for functional determinants, we have {(see eq.~(\ref{equ:detrelation2}) in appendix~\ref{app:constdet})}
\begin{equation}\label{eq:condet}
    {\text{Det}}_{\bar\xi}\,M_{\bar\xi}=\nu(\bar\xi)\text{Det}\, \tilde M_\kappa~,
\end{equation}
where 
\begin{equation}\label{eq:fluctopkappa}
    \tilde M_\kappa(x,y)=
    \left.\frac{\delta^2 \tilde S_\kappa}{\delta \phi(x) \delta \phi(y)}
    \right|_{\phi=\tilde\phi_\kappa}
    ,
\end{equation}
and $\nu(\bar\xi)$ is a real projection factor. 
It is given by (see eq.~(\ref{equ:nufunc})),
\begin{equation}
\label{eq:projectionfactor}
    \nu(\bar\xi)=\frac{\int d^4x \zeta(x)\psi(x)}{\int d^4x \zeta(x)^2},
\end{equation}
where
\begin{equation}
\label{equ:zetadef}
        \zeta(x)=\left.\frac{\delta\xi[\phi]}{\delta\phi(x)}\right|_{\phi=\phi_{\bar\xi}}
    =\left.\frac{d\mathcal{O}}{d\phi}
    \right|_{\phi=\phi_{\bar\xi}(x)},
\end{equation}
and $\psi(x)$ is a solution of the linear equation
\begin{equation}
\label{equ:psidef}
       \int d^4y\,  \tilde{M}_\kappa(x,y)\psi(y)=\zeta(x)~.
\end{equation}
We give a proof of this for finite vector spaces in appendix~\ref{app:constdet}.

Using eq.~(\ref{eq:condet}), we can write the vacuum decay rate (\ref{equ:constGamma}) as
\begin{equation}
    \label{equ:constGamma2}
    \Gamma\approx
    \KGdel{2}
    \int d\bar\xi\,\left(\frac{S[\phi^{\rm inst}_{\bar\xi}]}{2\pi}\right)^2
    \left|
        \frac{\nu(\bar\xi)\, {\overline{\text{Det}}} \tilde{M}_\kappa[\phi^{\rm inst}_{\bar\xi}]}
        {{\text{Det}} M[\phi^0]}
        \right|^{-1/2}e^{-\left(S_E[\phi^{\rm inst}_{\bar\xi}]-S_E[\phi^0]\right)}~,
\end{equation}
where both determinants are calculated in the space of all functions. In this paper, we do not need them because we only focus on the exponential factor in eq.~(\ref{equ:constGamma2}), but as we will discuss in section~\ref{sec:negativemodes}, eq.~(\ref{eq:condet}) also gives as a convenient way to determine the number of negative modes and therefore identify the relevant constrained instanton solutions.

\section{Scalar field theory} \label{sec:massphi}

\subsection{Massless vs massive $\phi^4$} \label{sec:fubini}

As an application of the formalism introduced in section~\ref{sec:constinst},
consider a theory of a single real scalar field in 4 dimensions with the potential
\begin{equation}\label{eq:massive-potential}
    V(\phi)=\frac{1}{2}m^2\phi^2-\frac{\lambda}{4!}\phi^4~,
\end{equation}
where $m^2 \geq0$ and $\lambda>0$. 
This theory is somewhat pathological in the sense that the potential is not bounded from below, but it has a metastable false vacuum located at the origin, $\phi^0=0$. We want to compute the lifetime of this false vacuum state.

The massless limit, $m^2=0$, of this theory is even more curious because the potential has no local minimum. Nevertheless, it has instanton solutions with a finite positive action \cite{Fubini:1976jm}.\footnote{The negative sign of the quartic term is crucial for this result. Clearly a $\phi^4$ theory with positive quartic term admits no tunnelling behaviour.} The Euclidean action for the massless theory is
\begin{equation}\label{eq:fubaction}
    S=\int d^4x \, \left[\frac{1}{2}(\partial_\mu\phi)^2 - \frac{\lambda}{4!}\phi^4\right]\,,
\end{equation}
and the equation of motion that follows is
\begin{equation}\label{eq:fubinieom}
    \partial^2\phi + \frac{\lambda}{6}\phi^3=0\,.
\end{equation}
Eq.~\eqref{eq:fubinieom} has an instanton solution 
\begin{equation}\label{eq:massless-instanton}
    \phi^{\rm inst}(x) = \frac{4\sqrt{3}\,\rho\,\lambda^{-1/2}}{\rho^2+x^2}\,,
\end{equation}
where $\rho$ is a free parameter arising as a result of the invariance of the action
(\ref{eq:fubaction}) under the scaling transformation
\begin{equation}\label{eq:scaltran}
    \phi \rightarrow a\phi(ax)\,,\quad a\in{\mathbb R}~,
\end{equation}
and can be interpreted as the size of the instanton. 
The action of the instanton is independent of this size parameter,
\begin{equation}\label{eq:mlessact}
    S[\phi^{\rm inst}]=\frac{16\pi^2}{\lambda}\,.
\end{equation}
We refer to eq.~\eqref{eq:massless-instanton} as the massless instanton. 

However, when $m^2>0$, no instanton solution exists.
The Euclidean action for this massive theory is 
\begin{equation}\label{eq:mveact}
    S=\int d^4x\,\left[\frac{1}{2}(\partial_\mu\phi)^2+\frac{1}{2}m^2\phi^2-\frac{\lambda}{4!}\phi^4\right]\,.
\end{equation}
Under the scaling transformation (\ref{eq:scaltran}), this changes as
\begin{equation}\label{eq:actscal}
    S \rightarrow 
    \int d^4x\,\left[\frac{1}{2}(\partial_\mu\phi)^2+a^{-2}\frac{1}{2}m^2\phi^2-\frac{\lambda}{4!}\phi^4\right]\,.
\end{equation}
Because $m^2\phi^2>0$ for any $\phi\neq 0$, the action of any non-trivial field configuration in this theory can always be lowered by making a scale transformation with $a>1$. Therefore the action (\ref{eq:mveact}) can have no stationary points other than the false vacuum, $\phi=0$.

To see how the constrained instanton approach circumvents the scaling argument, consider the modified action \eqref{eq:deformact} in the massive theory
\begin{equation}\label{eq:mphideformact}
    \tilde S_\kappa= \int d^4x\,\left[\frac{1}{2}(\partial_\mu\phi)^2+\frac{1}{2}m^2\phi^2-\frac{\lambda}{4!}\phi^4\right]+\kappa\int d^4x\,\mathcal{O}\,,
\end{equation}
with a constraint operator that scales with scaling dimension $d$ under the transformation (\ref{eq:scaltran}), i.e.,
$\mathcal{O}\rightarrow a^d \mathcal{O}$.
The modified action (\ref{eq:mphideformact}) then scales as
\begin{equation}
\label{equ:Skappascaling}
    \tilde S_\kappa\rightarrow 
    \int d^4x\,\left[\frac{1}{2}(\partial_\mu\phi)^2+a^{-2}\frac{1}{2}m^2\phi^2-\frac{\lambda}{4!}\phi^4\right]+a^{d-4}\,\kappa\int d^4x\,\mathcal{O}\,.
\end{equation}
For this to be a non-monotonic function of $a$, we must have $d\neq 2$ and $d\neq 4$. Moreover, demanding the action be stationary with respect to the scaling transformation \eqref{eq:scaltran} (at $a=1$) leads to the condition
\begin{equation}\label{eq:opcond}
    m^2\int d^4x \,\phi^2 = (d-4)\kappa\int d^4x\,\mathcal{O}(\phi) \,,
\end{equation}
which will be useful later. 

The simplest choice for the constraint operator is
\begin{equation}
\label{equ:constraintO}
    \mathcal{O}=\phi^d~.
\end{equation}
In sections~\ref{sec:phi3} and~\ref{sec:phi6} we will study the constrained instantons for two different constraint operators of this form --- $\phi^3$ and $\phi^6$ respectively.

\subsection{Numerical setup}\label{sec:numerics}

To find the constrained instantons, we need to solve the equation of motion following from the modified action \eqref{eq:mphideformact},
\begin{equation}
    \partial^2\phi-m^2\phi+\frac{\lambda}{6}\phi^3+\kappa\frac{\delta\xi}{\delta \phi}=0\,,
\end{equation}
for all possible values of the Lagrange multiplier $\kappa$. The problem can be simplified by noting that the theory has $O(4)$ symmetry and assuming that the solution has this same symmetry. As a consequence, the equations of motion can be written as a single ODE in the 4D radial variable $r=|x|$,
\begin{equation}\label{eq:eomODE}
    \phi''+\frac{3}{r}\phi'-m^2\phi+\frac{\lambda}{6}\phi^3+\kappa\frac{\delta \xi}{\delta \phi}=0\,,
\end{equation}
where the prime $(')$ denotes differentiation with respect to $r$. The boundary conditions are
\begin{equation}\label{eq:smallrBC}
    \phi'(0)=0~,
\end{equation}
corresponding to differentiability of the solution at the origin in four dimensions, and
\begin{equation}\label{eq:largerBC}
    \lim_{r\to\infty}\phi(r)=0~,
\end{equation}
which follows from the boundary conditions of eq.~(\ref{equ:GammaPI}).

We can deduce the behaviour of the constrained instanton solutions at small and large distances by considering the behaviour of eq.~\eqref{eq:eomODE} in those regimes. Due to the presence of the mass term which dominates the equation of motion at large distances, the constrained instanton solutions are expected to fall off exponentially for $r \gg m^{-1}$ which is in sharp contrast to to the {$1/r^2$} falloff of the massless instantons~\eqref{eq:massless-instanton}. Near the origin, $r \ll m^{-1}$, they are expected to resemble the massless instantons for some instanton size $\rho$.  Summarising, we expect our numerical results to fit to
\begin{equation}
    \phi(r) \approx 
    \begin{cases}
    \displaystyle
    \frac{4 \sqrt 3 \rho \lambda^{-1/2}}{\rho^2 + r^2}, & \text{if }r \ll m^{-1}~, \\
    \displaystyle
    A \frac{e^{- m r}}{r^{3/2}}, &\text{if } r \gg m^{-1}~,
    \end{cases}
    \label{eq:small-large-fits}
\end{equation}
for fitting parameters $A$ and $\rho$.

In general, the constrained equation of motion cannot be solved analytically, even for a simple choice of the constraint operator $\mathcal{O}$. Instead, we solve it numerically using the shooting method (see e.g. ref.~\cite{Press:2007ipz}), using the field value at the origin, $\phi(0)$, as the shooting parameter. Starting from an initial trial value, eq.~(\ref{eq:eomODE}) is solved as an initial value problem. Depending on whether the solution undershoots or overshoots the boundary condition (\ref{eq:largerBC}) at infinity, the initial value $\phi(0)$ is adjusted, and the process is iterated until eq.~(\ref{eq:largerBC}) is satisfied to a desired tolerance.

In practice, we solve the equation on a finite interval $r\in[R_{\min}, R_{\max}]$,
where $R_{\min}>0$ to avoid the singularity in eq.~(\ref{eq:eomODE}) but sufficiently small to fully capture the short-distance behaviour of the solution. Correspondingly, $R_{\max}$ has to be finite for a numerical solution but sufficiently large to allow accurate description of the long-distance tail. A detailed analysis of the impact of varying $R_{\min}$ and $R_{\max}$ on the accuracy of our results is presented in appendix~\ref{app:numcheck}.

\subsection{Negative modes}\label{sec:negativemodes}

As {discussed} in section~\ref{sec:constinst}, instantons are solutions with a single negative mode. Therefore, to identify the solutions we have found, we need to calculate the number of negative modes they have.

For ordinary instantons, counting the negative modes is straightforward.
Given a real scalar field theory with a generic Euclidean action 
\begin{equation}\label{eq:genact}
    S=\int d^4x \left(\frac{1}{2}(\partial_\mu\phi)^2 + V(\phi)\right),
\end{equation}
the fluctuation operator around a solution $\hat\phi$ is
\begin{equation}
    M(x,y)=
    \delta^4(x-y)\left(-\partial^2+V''\left(\hat\phi(x)\right)\right).
\end{equation}
Its eigenvalues $\Lambda$ can be found by solving the eigenvalue equation
\begin{equation}\label{eq:flucteq}
    \int d^4y\,M(x,y)
    \delta\phi(y)
    =\Lambda\delta\phi(x)~.
\end{equation}
For a spherically symmetric solution $\hat\phi$, we can separate the angular and radial equations. The solutions of the angular equation are four-dimensional spherical harmonics, labelled by an integer $\ell$. The radial equation then becomes
\begin{equation}
\label{eq:evaleq}
    \left(-\frac{d^2}{dr^2}-\frac{3}{r}\frac{d}{dr} 
    +\frac{\ell(\ell+2)}{r^2}+
{   V''\left(\hat\phi(r)\right)}
    \right)\delta\phi_{n\ell}(r)=\Lambda_{n\ell}\delta\phi_{n\ell}(r)~.
\end{equation}
This equation is of the Sturm-Liouville form, and therefore the $n$th eigenfunction has $n-1$ nodes.

For $\ell=1$, there is a zero mode $\delta\phi(r)=d\hat\phi(r)/dr$ associated with the broken translation invariance. If $\hat\phi$ is a monotonic function, then this eigenfunction has no nodes, and it is therefore the lowest one. This means that there can be no negative modes with $\ell=1$. 
{Because the $\ell$-dependent term is everywhere positive and monotonic in $\ell$, the eigenvalues satisfy $\Lambda_{n,\ell+1}\ge \Lambda_{n\ell}.$} Therefore, if there are any negative modes, they must have $\ell=0$. To determine how many there are, we solve eq.~(\ref{eq:evaleq}) with $\ell=0$ and $\Lambda_{n0}=0$ and count the nodes of the solution.

This same procedure does not work when a constraint is present, because the fluctuation operator $M_{\bar\xi}$, defined in the constrained function space $\mathcal{F}_{\bar\xi}$, is not a local differential operator and therefore the equivalent of eq.~(\ref{eq:flucteq}) would not be straightforward to solve. However, we can use eq.~(\ref{eq:condet}), which allows us to express its determinant in terms of $\tilde{M}_\kappa$ defined in eq.~(\ref{eq:fluctopkappa}), i.e.,
\begin{equation}
\label{equ:Mtilde}
 \tilde{M}_\kappa(x,y)
 =
  \delta^4(x-y)\Bigl(-\partial^2+V''\left(\phi_{\bar\xi}(x)\right)
  +\kappa\mathcal{O}''\left(\phi_{\bar\xi}(x)\right)\Bigr).
\end{equation}
As this is an \emph{unconstrained} fluctuation operator, defined in the space of all functions, it \emph{can} be treated using the procedure above.

For a finite interval $[R_{\rm min},R_{\rm max}]$, the operator $\tilde{M}_\kappa$ has a discrete spectrum of eigenvalues $\{\tilde\Lambda_{n\ell}\}$. Because the operator $M_{\bar\xi}$ can be obtained from $\tilde{M}_\kappa$ through an orthogonal projection in the space of functions (see appendix~\ref{app:constdet}), the Cauchy Interlacing Theorem \cite{doi:10.1137/1.9781611971170} shows that its eigenvalues $\{\Lambda_{n\ell}\}$ satisfy the inequality
\begin{equation}
\tilde\Lambda_{1,0}\leq\Lambda_{1,0}\leq\tilde\Lambda_{2,0}\leq\Lambda_{2,0}\leq\tilde\Lambda_{3,0}\le\ldots.
\end{equation}
Hence, if $\tilde{M}_\kappa$ has a single negative mode, i.e., $\tilde{\Lambda}_{1,0}<0<\tilde{\Lambda}_{2,0}$, then
the projected operator
$M_{\bar\xi}$ has either one negative mode or none.
And because the sign of the determinant tells whether the operator has an odd or even number of negative modes, eq.~(\ref{eq:condet}) shows that the operator $M_{\bar\xi}$ has one negative mode if $\nu(\bar\xi)>0$ and no negative modes of $\nu(\bar\xi)<0$, where 
$\nu(\bar\xi)$ is defined in eq.~(\ref{eq:projectionfactor}). Thus, the prescription for computing the number of negative modes around a constrained solution as follows. First, one computes the number of negative modes of $\tilde{M}_\kappa$ in eq.~\eqref{equ:Mtilde} using the standard approach described above. Then, one computes the projection factor $\nu(\bar\xi)$ (eq.~\eqref{eq:projectionfactor}) to determine whether any negative modes have been removed by the constraint.

In order to compute $\nu(\bar\xi)$ numerically, given a constrained instanton configuration $\phi_{\bar\xi}$, we first find the function $\zeta(x)$ defined in eq.~\eqref{equ:zetadef}, 
and then obtain the auxiliary function $\psi(x)$ by solving eq.~(\ref{equ:psidef}), which has the explicit form
\begin{equation}
    \label{eq:psieom}
    \Bigl[
    -\partial^2+V''\left(\phi_{\bar\xi}(x)\right)+\kappa{\cal O}''\left(\phi_{\bar\xi}(x)\right)
    \Bigr]
    \psi(x)
    =\zeta(x)~.
\end{equation}
We can choose the spherically symmetric solution, so this becomes an ordinary inhomogeneous differential equation in the radial coordinate $r$, with boundary conditions $\psi'(0)=\psi(\infty)=0$. Using the the function $\phi_{\bar\xi}$ obtained previously, we solve eq.~(\ref{eq:psieom}) on the same interval $[R_{\rm min},R_{\rm max}]$ using the shooting algorithm described in section~\ref{sec:numerics}, with $\psi(R_{\rm min})$ as the shooting parameter. The projection factor $\nu(\bar\xi)$ is then obtained by substituting this solution to eq.~(\ref{eq:projectionfactor}).

\section{Numerical results\label{sec:results}}

\subsection{Parameters}
\label{sec:params}
For a constraint of the form (\ref{equ:constraintO}), the modified action becomes
\begin{equation}\label{eq:deformact2}
    \tilde S_\kappa = \int d^4 x \left[ \frac{1}{2}(\partial_{\mu}\phi)^2 + \frac{1}{2} m^2 \phi^2  -\frac{\lambda}{4!}\phi^4 +\kappa\phi^d \right]~.
\end{equation}
For the numerical implementation, it is convenient to define dimensionless variables
\begin{equation}
\label{equ:dimlessvar}
    X\equiv mx~,\quad \Phi\equiv\frac{\lambda^{1/2}}{m}\phi~,\quad
    K_d\equiv \frac{m^{d-4}}{\lambda^{d/2-1}}\kappa~.
\end{equation}
In terms of them, the action (\ref{eq:deformact2}) becomes
\begin{equation}
    \tilde S_\kappa=\frac{1}{\lambda}
    \int d^4 X \left[ \frac{1}{2}\left(\frac{\partial\Phi}{\partial X^\mu}\right)^2 + \frac{1}{2} \Phi^2  -\frac{1}{4!}\Phi^4 +K_d\Phi^d \right]~.
\end{equation}
This shows that for a given $d$, the solutions only depend on one parameter $K_d$. For a fixed $K_d$, the product $\lambda \tilde{S}_\kappa$ is independent of both $\lambda$ and $m$, too. We use the dimensionless variables (\ref{equ:dimlessvar}) internally for the numerical calculations but present the results in terms of the physical, dimensionful variables.

\subsection{$\phi^3$ constraint}\label{sec:phi3}

We first consider a constraint operator that is cubic in the field $\phi$
\begin{equation}
    \mathcal{O}=\phi^3~.
\end{equation}
This results in the modified action 
\begin{equation}\label{eq:phi3deformact}
    \tilde S_\kappa = \int d^4 x \left[ \frac{1}{2}(\partial_{\mu}\phi)^2 + \frac{1}{2} m^2 \phi^2  -\frac{\lambda}{4!}\phi^4 
    +\kappa \phi^3 \right]~.
\end{equation}

As shown in section~\ref{sec:params}, the solutions depend only on the dimensionless parameter $K_3 \equiv \kappa/(m\lambda^{1/2})$.
We choose $\kappa<0$ to obtain positive solutions and therefore $\bar\xi>0$ (according to eq.~\eqref{eq:opcond}). Because of the $\mathbb{Z}_2$ symmetry of the original action (\ref{eq:mveact}), the solutions for $\kappa>0$ are given by simply changing the sign, $\phi\rightarrow -\phi$.

\begin{figure}[t]
     \centering
     \includegraphics[width=0.7\linewidth]{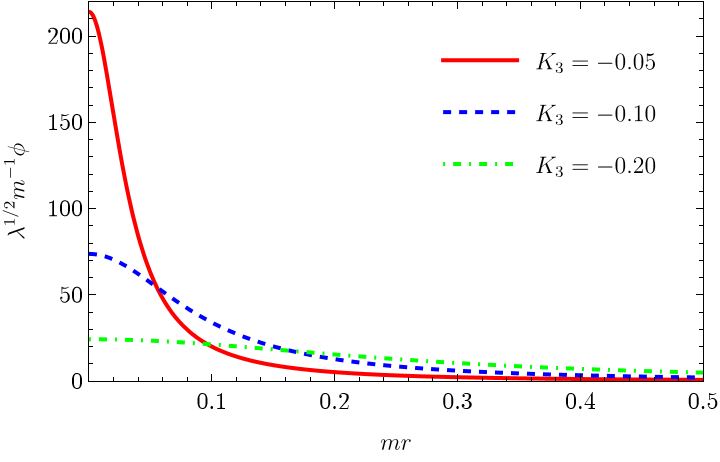}
     \caption{
     Solutions for the $\phi^3$ constraint for several values of $K_3\equiv \kappa/(m\lambda^{1/2})$.}
     \label{fig:phi3CIs}
 \end{figure}

As discussed in section~\ref{sec:numerics}, we set up the numerical problem on a finite interval with $R_{\max}=50m^{-1}$ for all investigated values of $K_3\in[-2,-10^{-5}]$.  This ensured that it was always at least one order of magnitude larger than the instanton size. The lower boundary $R_\mathrm{min}$ was set according to
\begin{equation}
    R_\mathrm{min} = \begin{cases}
        10^{-7}m^{-1}~,   & K_3 < -9 \times 10^{-4}~,\\
        10^{-10}m^{-1}~,  & \mathrm{otherwise}~.
    \end{cases}
\end{equation}
The different choices of $R_{\min}$ are motivated by the need to capture the correct behaviour of the solutions near the origin. In the case of very small $|\kappa|$, we found that the field profiles become extremely narrow and it is necessary to decrease $R_{\min}$ accordingly, to ensure that it is always at least three orders of magnitude smaller than the instanton size.  The shooting algorithm was iterated until the boundary conditions were satisfied to within an absolute uncertainty of $\pm2\times10^{-14}$. In this setup, we found 790 solutions, some of which are shown in figure~\ref{fig:phi3CIs}.

\begin{figure}[t]
    \centering
    \includegraphics[width=0.45\textwidth]{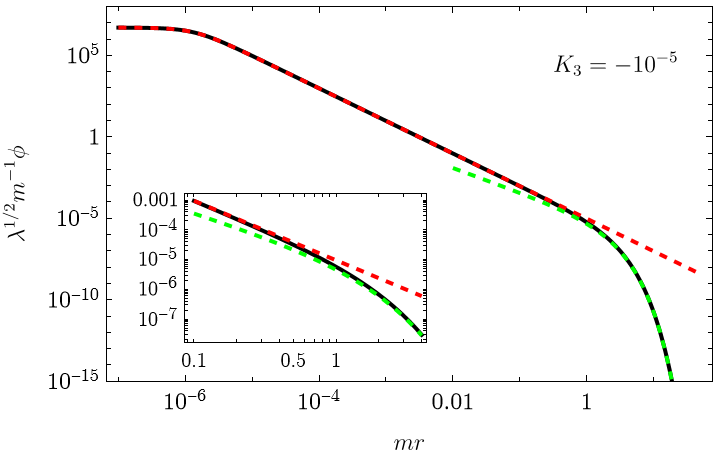}
    \qquad
    \includegraphics[width=0.45\textwidth]{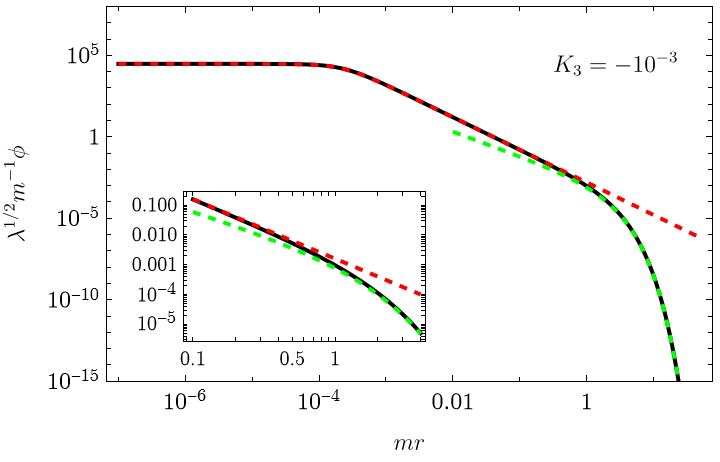}
    \qquad
    \includegraphics[width=0.45\textwidth]{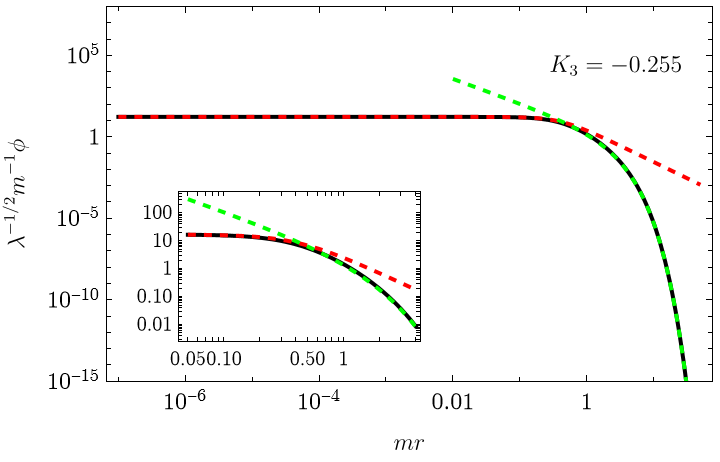}
    \qquad
    \includegraphics[width=0.45\textwidth]{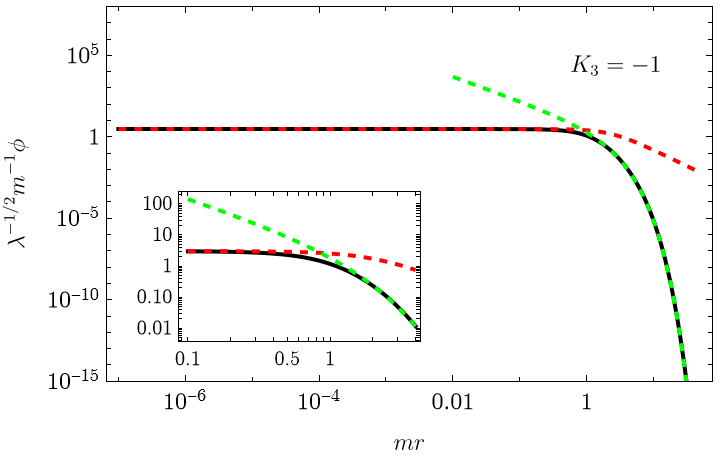}
    \caption{ Constrained instantons for the $\phi^3$ constraint at different values of $K_3\equiv \kappa/(m\lambda^{1/2})$. 
    The red and green dashed lines show the short and long distance fits given by eq.~\eqref{eq:small-large-fits}}
    \label{fig:phi3CIapprox}
\end{figure}

In figure~\ref{fig:phi3CIapprox} , we compare solutions for selected values of $\kappa$ to the analytic approximations (\ref{eq:small-large-fits}) for small and large radii. We found that for larger solutions, the range of validity of the massless approximation decreases, as expected. We also found the expected exponential falloff behaviour for all of our solutions.

\begin{figure}[t]
     \centering
     \includegraphics[width=0.45\textwidth]{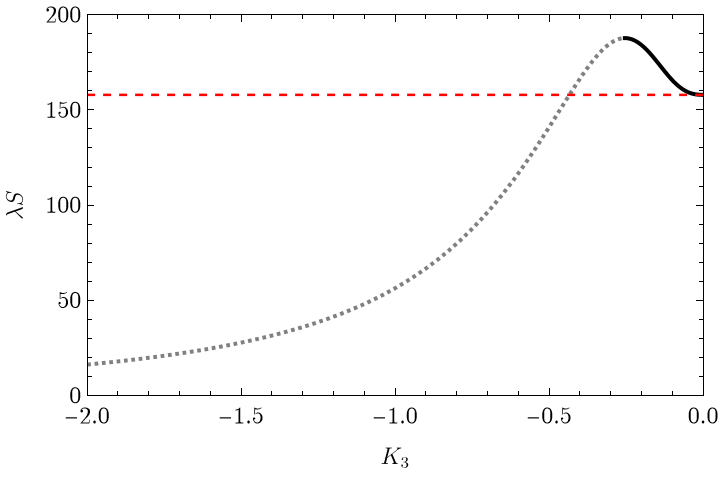}
     \qquad
     \includegraphics[width=0.45\textwidth]{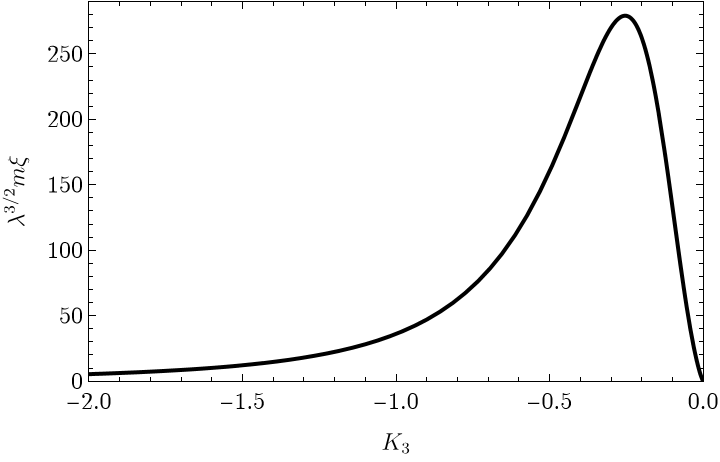}
     \caption{Left: The action $S$ of our solutions as a function of $K_3\equiv \kappa/(m\lambda^{1/2})$ for the $\phi^3$ constraint. The dashed red line denotes the value of the massless instanton action \eqref{eq:mlessact}. The solid black line represents the solutions that contribute to the tunnelling rate, while the grey dashed line shows the ones that do not. Right: The constraint $\xi$ as a function of $K_3$ for the same solutions.}
     \label{fig:phi3SofK}
 \end{figure}

In the left panel of figure~\ref{fig:phi3SofK} we show the action $S$, defined in eq.~\eqref{eq:mveact}, of our solutions as a function of the Lagrange multiplier $\kappa$. As $\kappa$ approaches 0 the action approaches the massless instanton action (  eq.~\eqref{eq:mlessact}) from above, with the difference between the two $\Delta S\approx 1.17\times10^{-8}$ at $K_3=-10^{-5}$. As $\kappa$ decreases, the action grows until it reaches a maximum value $\lambda S_{\max}\approx 187.723$ at $K_\mathrm{3crit} \equiv \kappa_\mathrm{crit}/(m\lambda^{1/2}) \approx -0.255$. It subsequently begins to monotonically decrease. The action becomes smaller than the massless instanton action eq.~\eqref{eq:mlessact} around $K_\mathrm{3} \approx -0.435$, and continues to decrease, while remaining positive, over the rest of the investigated range of $\kappa$. However, since the action eq.~\eqref{eq:mveact}, is not positive definite, it is possible that the instanton action becomes negative outside of the investigated range of $\kappa$.

We then computed the value of the constraint $\bar \xi$ by evaluating eq.~\eqref{eq:constraint} on the constrained instanton solution. The value of the constraint as a function of the Lagrange multiplier $\kappa$ is shown in the right panel of figure~\ref{fig:phi3SofK}. We see that the constraint is not a monotonic function of the Lagrange multiplier $\kappa$.
Instead, like $S$ it has a maximum $\bar\xi_{\max}$ at $\kappa_{\rm crit}$, as required by eqs.~\eqref{eq:deformact} and \eqref{eq:legendrexi}. There are two critical consequences of the non-monotonicity of $\bar\xi(\kappa)$. The first one is that the integral over $\bar\xi$ in eq.~\eqref{equ:constGamma} 
is over a finite range $[0,\bar\xi_{\max}]$. The other implication is that there are two branches of solutions, that is, for each value of $\bar\xi$ there are two distinct solutions. We will discuss their interpretation shortly.

\begin{figure}[t]
    \centering
    \includegraphics[width=0.7\linewidth]{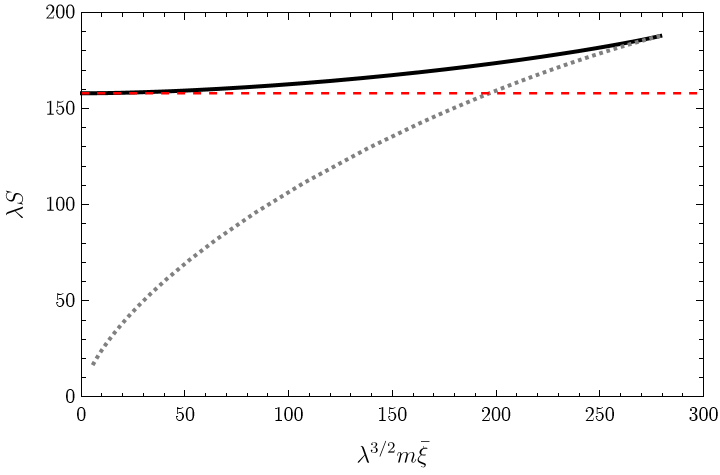}
    \caption{The action as a function of the constraint $\bar\xi$ for the $\phi^3$ constraint. The black solid line corresponds to the solutions that contribute to the tunnelling rate. The dashed red line shows the massless instanton action.}
    \label{fig:phi3SofXi}
\end{figure}

Having computed the constraint, we can re-express all physically meaningful quantities in terms of $\bar\xi$ rather than $\kappa$. In figure~\ref{fig:phi3SofXi} we plot the action \eqref{eq:mveact} as a function of the constraint. The two-branch structure is evident. The upper branch corresponds to $\kappa > \kappa_{\rm crit}$, while the lower branch corresponds to $\kappa<\kappa_{crit}$. On both branches, the action is a monotonically increasing function of $\bar\xi$, but the limits as $\bar\xi \to 0$ are different. For the upper branch, the $\bar\xi \to 0$ limit corresponds to the $\kappa \to 0$ limit and the action approaches the massless instanton action, eq.~\eqref{eq:mlessact}. For the lower branch, the same limit in $\bar\xi$ corresponds to decreasing $\kappa$. In this scenario, the action appears to approach 0. The two branches meet at a sharp cusp at the maximum value of the constraint, $\bar\xi_{\max}$. In order to see that the meeting point is a cusp rather than a smooth peak, we use eq.~\eqref{eq:legendrekap}. Because $\kappa<0$, $dS/d\bar\xi$ must \emph{always} be positive. However, a smooth transition from the lower to the upper branch would require $dS/d\bar\xi$ to change sign. This is not allowed. Therefore, the transition between the two branches cannot be smooth. 

To interpret the two branches of solutions, we studied the number of negative modes in the constrained fluctuation spectrum around the solutions for all investigated values of $\kappa$, following the procedure described in section~\ref{sec:negativemodes}. Unsurprisingly, we found that there is precisely one negative eigenvalue in the spectrum of the unconstrained operator $\tilde M_\kappa[\phi]$ (eq.~\eqref{eq:fluctopkappa}) for all values of $\kappa$. Therefore, depending on the sign of the projection factor $\nu$ (eq.~\eqref{eq:projectionfactor}), the constrained spectrum can contain up to one negative eigenvalue. 

\begin{figure}[t]
    \centering
    \includegraphics[width=0.7\linewidth]{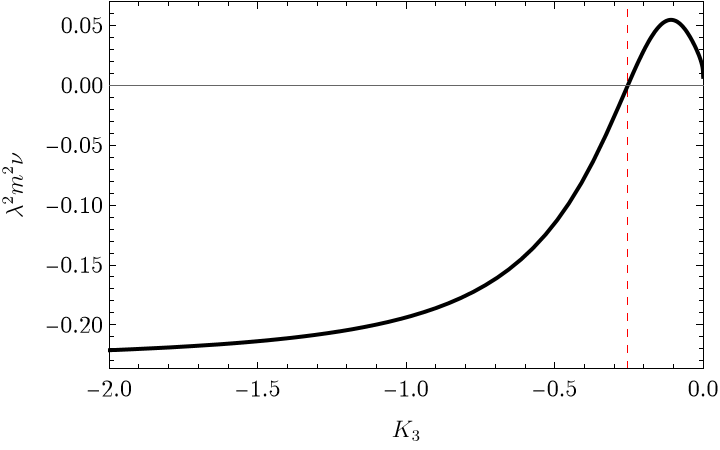}
    \caption{The projection prefactor $\nu$ as a function of $K_3\equiv \kappa/(m\lambda^{1/2})$ for the $\phi^3$ constraint. The red dashed line marks $\kappa = \kappa_\mathrm{crit}$.}
    \label{fig:phi3projectionfactor}
\end{figure}

The projection factor plotted as a function of $\kappa$ is shown in figure~\ref{fig:phi3projectionfactor}. We see that $\nu$ is positive in the upper range of $\kappa$, and turns negative at increasingly negative $\kappa$.  The zero crossing appears to precisely coincide (up to our numerical resolution in $\kappa$) with $\kappa = \kappa_\mathrm{crit}$. This tells us that the change in the number of negative modes happens right at the cusp in figure~\ref{fig:phi3SofXi}, with the upper (small $|\kappa|$) branch retaining the single negative mode from the unconstrained fluctuation operator $\tilde M_\kappa$, and the lower (large $|\kappa|$) branch having the mode removed by the projection factor $\nu$. The lack of a negative mode in the spectrum of the solutions on the lower branch implies that these solutions are not instantons. Instead, they are minima $\phi^0_{\bar\xi}$ of the action and contribute only to the first integral in eq.~\eqref{eq:fullconPI}. The solutions on the upper branch all have a single negative mode in their fluctuation spectrum --- these are the constrained instantons we aimed to find.

\subsection{$\phi^6$ constraint}
\label{sec:phi6}

We now consider another constraint operator
\begin{equation}
    \mathcal{O}=\phi^6.
\end{equation}
The modified action is
\begin{equation}\label{eq:phi6deformact}
    S = \int d^4 x \, \left[ \frac{1}{2}(\partial_{\mu}\phi)^2 + \frac{1}{2}m^2\phi^2 - \frac{\lambda}{4!}\phi^4 
    {+\kappa\phi^6}
    \right]~.
\end{equation}
For this choice of the constraint operator, the Lagrange multiplier $\kappa$ can only take values in a finite range,
\begin{equation}
    0 < \kappa < \kappa_{\max} = \frac{\lambda^2}{2\cdot (4!)^2 m^2}\approx 8.7\times 10^{-4}\frac{\lambda^2}{m^2}.
\end{equation}
The first inequality is needed for the modified potential to have a minimum at $\phi\ne 0$, and the second for it to be the global minimum.

\begin{figure}[t]
    \centering
    \includegraphics[width=0.7\linewidth]{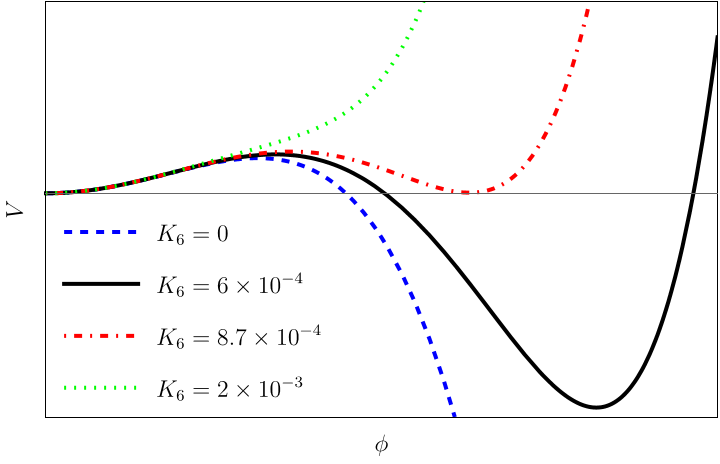}
    \caption{Variation in the shape of the modified potential with the $\phi^6$ constraint for different values of $K_6\equiv (m^2/\lambda^2)\kappa$.}
    \label{fig:phi6pot}
\end{figure}

\begin{figure}[t]
    \centering
    \includegraphics[width=0.7\linewidth]{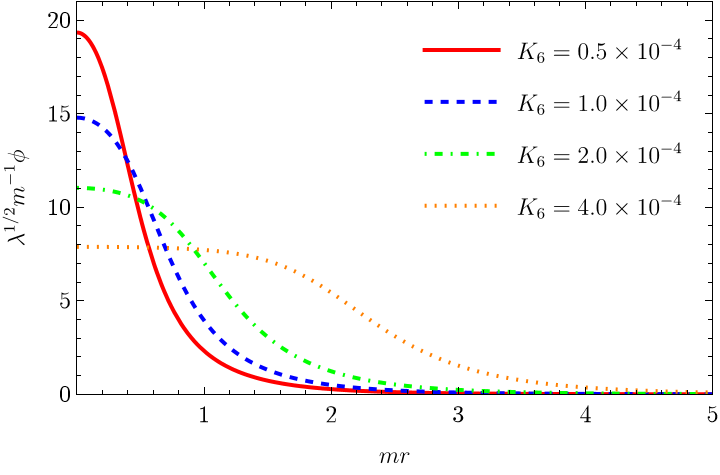}
    \caption{Solutions for several values of $K_6\equiv (m^2/\lambda^2)\kappa$ with the $\phi^6$ constraint.}
    \label{fig:phi6CIs}
\end{figure}

We set up the problem in a finite simulation box with $R_\mathrm{min} = 10^{-7}m^{-1}$ for the entire investigated range of $K_6\equiv(m^2/\lambda^2)\kappa \in [10^{-11}, 7.68\times 10^{-4}]$.
This ensured that $R_\mathrm{min}$ was always at least 4 orders of magnitude smaller than the instanton size.  The upper boundary $R_\mathrm{max}$ was chosen to always be at least three times the instanton size, which required a different choice depending on the size of $\kappa$:
\begin{equation}
    R_\mathrm{max} = \begin{cases}
        30m^{-1}~, & 10^{-11} \leq K_6 \leq 6 \times 10^{-7}~, \\
        50m^{-1}~, & \mathrm{otherwise}~.
    \end{cases}
\end{equation}
The instanton size diverges as $\kappa\rightarrow \kappa_\mathrm{max}$, making it impossible to cover the whole range of $\kappa$ with a finite simulation box size.  This limit will be studied in detail in an upcoming work~\cite{asymptotic-limits}. Some of the solutions for different values of $\kappa$ can be seen in figure~\ref{fig:phi6CIs}.

\begin{figure}[t]
    \centering
    \includegraphics[width=0.45\textwidth]{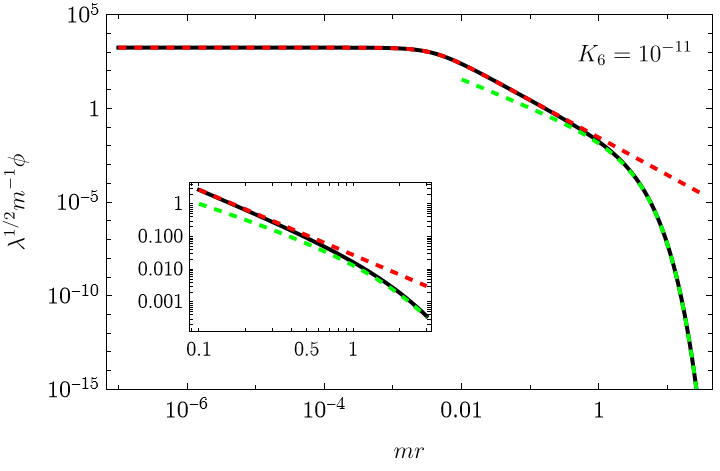}
    \qquad    
    \includegraphics[width=0.45\textwidth]{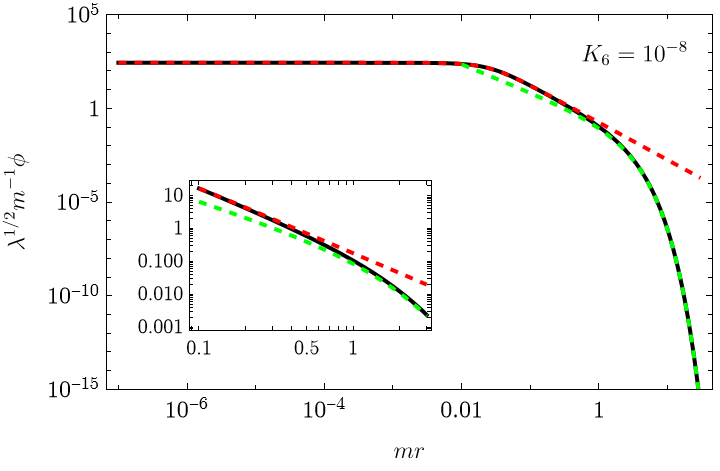}
    \qquad    
    \includegraphics[width=0.45\textwidth]{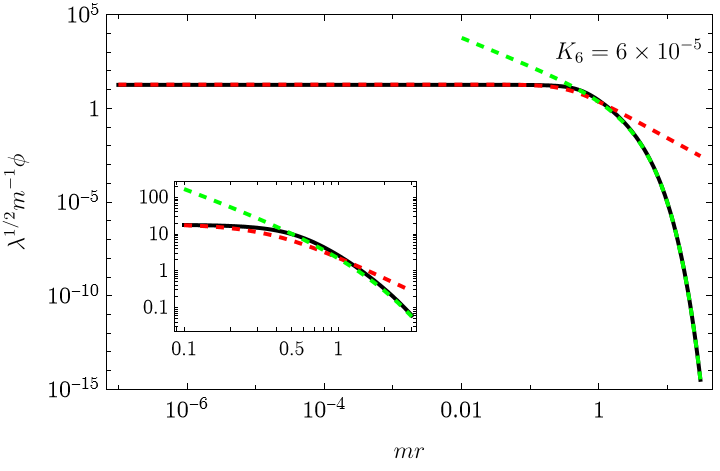}
    \qquad
    \includegraphics[width=0.45\textwidth]{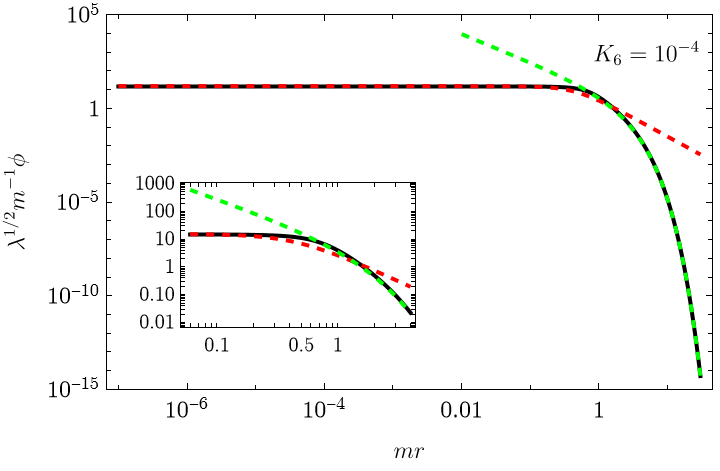}
    \qquad
    \caption{Solutions for different values of $K_6\equiv (m^2/\lambda^2)\kappa$ with the $\phi^6$ constraint, together with the short-range massless instanton fits (dashed red) and long-range exponential fits (dashed green) (eq.~\eqref{eq:small-large-fits}).}
    \label{fig:phi6CIsapprox}
\end{figure}

\begin{figure}[t]
     \centering
      \includegraphics[width=0.45\textwidth]{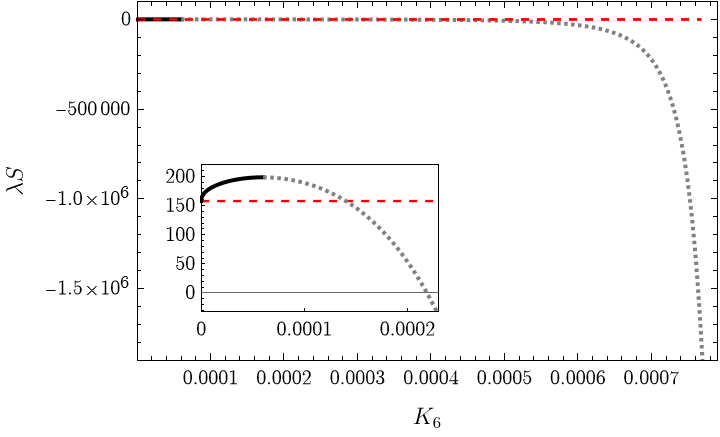}
     \qquad
     \includegraphics[width=0.45\linewidth]{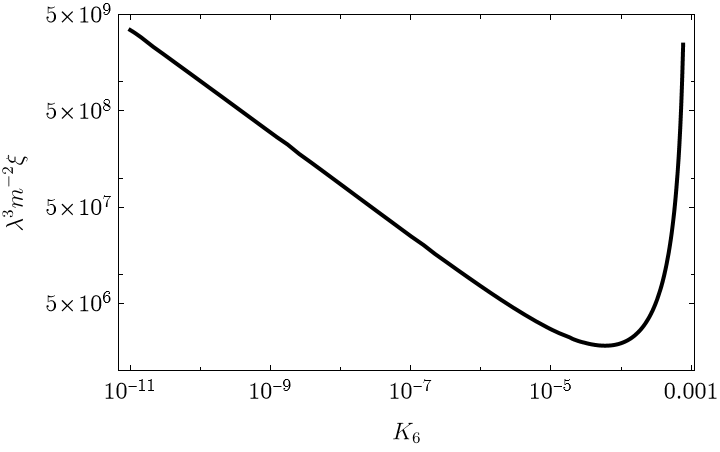}
     \caption{Action and constraint values for the $\phi^6$ constraint. Left: The action as a function of $K_6\equiv (m^2/\lambda^2)\kappa$. The red line denotes the value of the corresponding massless instanton action. The solid black line indicates the constrained instantons, that contribute to the vacuum decay rate, while the grey dashed line indicates those solutions that do not. Right: The constraint as a function of $K_6$.}
     \label{fig:phi6SofK}
 \end{figure}

As in the previous section, we tested the asymptotic behaviour of the solutions. Again, we found a good agreement with the analytic predictions, as seen in figure~\ref{fig:phi6CIsapprox}.
The plot of the action~\eqref{eq:mveact} as a function of $\kappa$ is shown in the left-hand-side panel of figure~\ref{fig:phi6SofK}. Initially it appears that the behaviour of the action is quite different from what we observed for the cubic constraint. For example, the action becomes negative within the investigated range of $\kappa$, at around $K_6 \approx 2.2 \times 10^{-4}$. As in the cubic case, the action monotonically approaches the massless instanton action from above in the limit of $\kappa \to 0$, with the difference between the two reaching $\Delta S\approx 0.038$ at $K_6=10^{-11}$. It then reaches a maximum value of $\lambda S_{\max}=198.754$ at a finite value of $\kappa$, here $K_{6 \rm crit}  \equiv (m^2/\lambda^2)\kappa_\mathrm{crit} \approx 6 \times 10^{-5}$, after which it decreases.

The constraint as a function of $\kappa$ is shown in the right-hand-side panel of figure~\ref{fig:phi6SofK}. Just as in the $\phi^3$ case, we see that $\xi(\kappa)$ is not monotonic. However, this time the constraint has a minimum rather than a maximum at $\kappa_{\rm crit}$. There are again two important conclusions to be drawn. The first regards the limits of integration over $\bar\xi$ in eq.~\eqref{equ:constGamma}. Since the constraint has a minimum, rather than a maximum, the lower limit is no longer at $\bar\xi=0$ but rather at the minimum value of $\bar\xi$. As for the upper limit, it is possible that it too is finite, although our numerical results are inconclusive on this matter and would be an interesting topic for future investigation. The second conclusion is again analogous to the one we made in the cubic case: There are two distinct solutions for each value of the constraint $\bar\xi$.\footnote{Strictly speaking there are four solutions to the $\phi^6$ problem, as can be seen by taking $\phi \to - \phi$.  The solutions are otherwise identical, so they may be accounted for by including a factor of 2 in the path integral.}

\begin{figure}[t]
    \centering
    \includegraphics[width=0.7\linewidth]{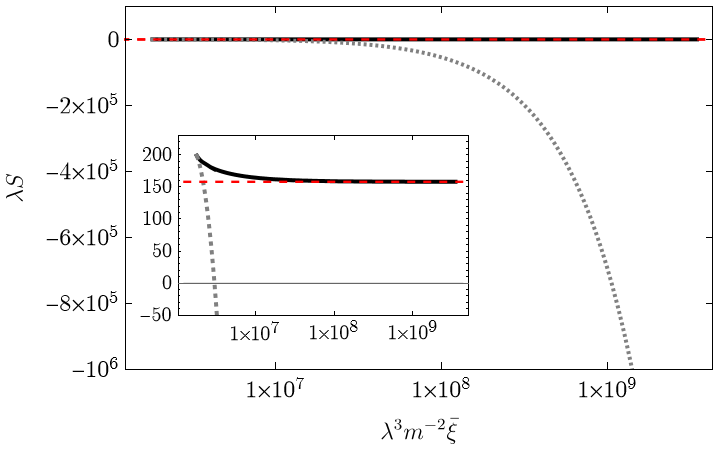}
    \caption{Action as a function of the constraint value for the $\phi^6$ constraint. The red line represents the corresponding value of the massless instanton action. The black solid line corresponds to the configurations that contribute to the tunnelling rate, while the grey dashed line denotes those that do not.}
    \label{fig:phi6SofXi}
\end{figure}

Having computed the constraint, we can re-express the action as a function of $\bar\xi$. This is shown in figure~\ref{fig:phi6SofXi}. Again, the two-branch structure is evident, but the action is now a monotonically decreasing function of $\bar\xi$. Again, the large $\bar\xi$ limits on the two branches correspond to different limits in $\kappa$. On the upper branch, the large $\bar\xi$ limit corresponds to $\kappa \to 0$. In this limit, the action approaches the massless instanton action. On the lower branch, the same limit corresponds to the $\kappa\to\kappa_{\max}$ limit, and the action decreases rapidly. On the other end, the branches again meet at a sharp cusp at $\bar\xi_{\min}\approx1.8\times10^6$.\footnote{For an argument that the meeting point is a cusp see section~\ref{sec:phi3}.}

\begin{figure}[t]
    \centering
    \includegraphics[width=0.7\linewidth]{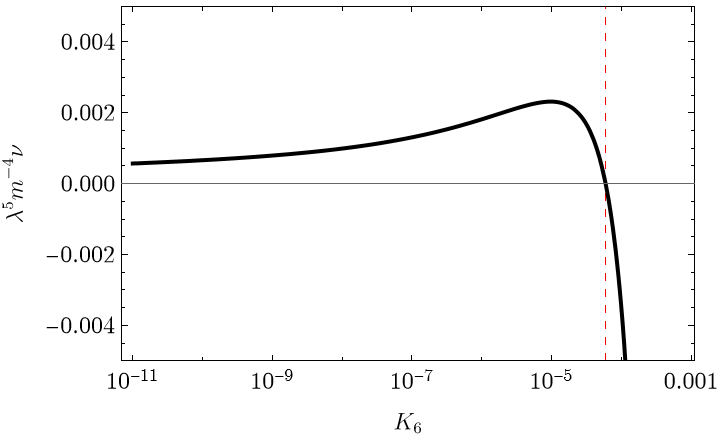}
    \caption{ 
    The projection prefactor $\nu$ as a function of $K_6\equiv (m^2/\lambda^2)\kappa$ for the $\phi^6$ constraint. The red dashed line marks $\kappa=\kappa_{\rm crit}$.}
    \label{fig:phi6nu}
\end{figure}

We again studied the number of negative modes in the constrained fluctuation spectrum around all of the constrained solutions. As for the case of the cubic constraint, we found that the unconstrained operator $\tilde M_\kappa[\phi]$ (eq.~\eqref{eq:fluctopkappa}) has exactly one negative eigenvalue in its fluctuation spectrum for all investigated values of $\kappa$. The number of negative modes in the constrained spectrum is then determined by the sign of the projection factor $\nu$ (eq.~\eqref{eq:projectionfactor}), which is shown as a function of $\kappa$  in figure~\ref{fig:phi6nu}. We see that $\nu$ becomes negative for $\kappa > \kappa_\mathrm{crit}$ (again, up to numerical resolution in $\kappa$). This implies that the solutions on the lower branch of the $S(\bar\xi)$ curve have no negative modes and are therefore minima $\phi^0_{\bar\xi}$ of the action, contributing only to the first integral in eq.~\eqref{eq:fullconPI}, while the top branch consists of the constrained instantons $\phi^{\rm inst}_{\bar\xi}$.

\section{Discussion and conclusions}\label{sec:concl}

In this work we have developed an explicit method of computing the vacuum decay rate in quantum field theory using constrained instantons, based upon earlier perturbative work by Affleck~\cite{Affleck:1980mp}.  
We applied this method to the massive $\phi^4$ theory with a negative self-coupling, considering two different constraint functions, $\phi^3$ and $\phi^6$.

For comparison with Affleck's original work~\cite{Affleck:1980mp}, the constraint values can be loosely mapped to the massless instanton size $\rho$, as $\rho\sim|\bar\xi|$ for $\phi^3$ and $\rho\sim \bar\xi^{-1/2}$ for $\phi^6$.
For both types of constraint we found solutions only for a subset of all possible values of the constraint $\bar\xi$. For $\phi^3$ this is a finite range around zero, and for $\phi^6$ the constraint $\bar\xi$ needs to be larger than a certain positive minimum value. In both cases these correspond to small instanton ``sizes'', in line with expectations~\cite{Affleck:1980mp}.
We do not expect that there are other relevant solutions outside these ranges, but we cannot rule it out completely based on our results.

For both constraints, we found a two-branch structure of solutions, shown in figures~\ref{fig:phi3SofXi} and \ref{fig:phi6SofXi}. By counting the negative modes we identified the lower branch as consisting of the minima of the action subject to the constraint, and the upper branch as the constrained instanton solutions. Only the latter contribute to the the vacuum decay rate (\ref{equ:constGamma2}). In both cases, the constrained instantons correspond to the small absolute values of the Lagrange multiplier $\kappa.$

Focusing on the constrained instanton solutions (i.e. the upper branch) we can see that the constrained instanton action is everywhere higher than that of the massless instanton. It seems to approach the massless instanton action in the limit $\bar\xi \to 0$ for the $\phi^3$ constraint and $\bar\xi \to \infty$ for the $\phi^6$ constraint, which both correspond to the $\kappa \to 0$ limit. It then monotonically increases while remaining finite as $\bar\xi$ approaches the end of the allowed range, with a different maximum value for the two constraints. It is generally well approximated by the massless instanton action, and the actual solutions shown in figures~\ref{fig:phi3CIapprox} and~\ref{fig:phi6CIsapprox} also resemble the massless instanton solutions for small $|\kappa|$,
which suggests that some kind of perturbative approach should be reasonably accurate. We will study this in more detail in a future work~\cite{asymptotic-limits}.

The vacuum decay rate can be obtained from eq.~(\ref{equ:constGamma2}) by integrating over the constraint $\bar\xi$.In the limit where the action approaches the massless instanton action, the exponential factor in the integrand is largest. However, in order to compute the rate accurately, one must calculate the functional determinants. This calculation and therefore the integration has to be carried out numerically and is left for future work. As a physical quantity, the rate should be independent of the choice of the constraint function, and therefore comparing the results obtained using the two constraints will provide a good indicator of the accuracy of the approximations involved in our method.

Even though the focus of this paper has been on the scalar theory, the constrained instanton method we have developed is applicable to other theories as well. 
It is straightforward to generalise it to the case of electroweak vacuum metastability~\cite{Degrassi:2012ry,Markkanen:2018pdo}, as the Higgs potential behaves approximately as eq.~(\ref{eq:fubaction}) for a wide range of values.
It would be interesting to apply it to the case of additional non-renormalisable operators, which have been found to have a significant effect in the conventional instanton approximation~\cite{Branchina:2014rva}.
Our constrained instanton method also provides a way to compute the rate of baryon number violation in the Standard Model. While it is expected to be far too low to be presently observable, it is nevertheless an important quantity characterising the fundamental laws of nature. We aim to address these questions in future works.\\

{\small
{\bf Acknowledgments}
The authors are grateful to José Ram\'on Espinosa, Oliver Gould, and Tanmay Vachaspati for helpful conversations.  A.R.\ and B.E.\ were supported by STFC Consolidated Grants ST/T000791/1 and ST/X000575/1, and B.E.\ also by Simons Investigator award 690508. K.G. was supported by the STFC DTP research studentship grant ST/X508433/1.}

\appendix

\section{Constrained functional determinant}
\label{app:constdet}

In this appendix we derive eqs.~\eqref{eq:condet} and~\eqref{eq:projectionfactor}
for finite-dimensional vector spaces. We believe that they are also valid for function spaces but do not have a rigorous proof. 

Consider a finite-dimensional vector space $V$, $\dim V=N$ with coordinates $\{x_i\}$, $i=1,2,...N$ and an action function $S(\vec x)$ defined on $V$. We are interested in a saddle point $\hat x$ {of $S(\vec{x})$} subject to the constraint
\begin{equation}
\label{eq:constr}
    \xi(\vec x)=\bar \xi~,
\end{equation}
where $\xi$ is a nonlinear function of $\vec x$ {and corresponds to the functional defined in eq.~(\ref{eq:constraint})}. This constraint defines an $(N-1)$-dimensional hypersurface $\mathcal{F}_{\bar\xi}\subset V$, here called the constraint surface. We parameterise the points on the constraint surface with a new set of coordinates $\{y_I\}$, $I=1,2,..., N-1$.

In terms of these coordinates, the saddle point $\hat{x}$ satisfies the equation
\begin{equation}
    \left.\frac{\partial S}{\partial y_I}\right\vert_{\hat x}=\left.\frac{\partial S}{\partial x_i}\right\vert_{\hat x}\left.\frac{\partial x_i}{\partial y_I}\right\vert_{\hat x}=0~.
\end{equation}
To compute the vacuum decay rate using eq.~(\ref{equ:constGamma}), we need the determinant of the $(N-1)\times(N-1)$ constrained Hessian matrix $M_{\bar\xi}$ around this saddle point. This matrix has elements
\begin{equation}\label{eq:secderiv}
M_{\bar\xi,IJ}\equiv
    \left.\frac{\partial^2S}{\partial y_I \partial y_J}\right\vert_{\hat x}=\left.\frac{\partial^2 S}{\partial x_i \partial x_j}\right\vert_{\hat x}\left.\frac{\partial x_i}{\partial y_I}\right\vert_{\hat x}\left.\frac{\partial x_j}{\partial y_J}\right\vert_{\hat x}+\left.\frac{\partial S}{\partial x_i}\right\vert_{\hat x}\left.\frac{\partial^2 x_i}{\partial y_I \partial y_J}\right\vert_{\hat x}~.
\end{equation}
The first term on the right hand side is just the linear projection of the unconstrained Hessian matrix to the tangent space of the constraint surface. The second term is present only for non-linear constraints. In a finite-dimensional vector space, it can be computed by an explicit construction of the coordinates $y_I$, but in a space of functions it is less straightforward, and therefore we take a different route.

We start by defining the modified action analogous to eq.~(\ref{eq:deformact}),
\begin{equation}
    \tilde{S}_\kappa(\vec{x})=S(\vec{x})+\kappa\xi(\vec{x}),
\end{equation}
with the Lagrange multiplier $\kappa$ chosen so that
\begin{equation}
\label{equ:dStilde}
{\left.\frac{\partial\tilde{S}_\kappa}{\partial x_i}\right|_{\hat{x}}=}
    \left.\frac{\partial (S+\kappa \xi)}{\partial x_i}\right\vert_{\hat x}=0~.
\end{equation}

For this function, we have
\begin{equation}\label{eq:secderiv2}
    \left.\frac{\partial^2\tilde{S}_\kappa}{\partial y_I \partial y_J}\right\vert_{\hat x}
    =
    \left.\frac{\partial^2 \tilde{S}_\kappa}{\partial x_i \partial x_j}\right\vert_{\hat x}\left.\frac{\partial x_i}{\partial y_I}\right\vert_{\hat x}\left.\frac{\partial x_j}{\partial y_J}\right\vert_{\hat x}
    +\left.\frac{\partial \tilde{S}_\kappa}{\partial x_i}\right\vert_{\hat x}\left.\frac{\partial^2 x_i}{\partial y_I \partial y_J}\right\vert_{\hat x}
    =    
    \tilde{M}_{\kappa,ij}
\left.\frac{\partial x_i}{\partial y_I}\right\vert_{\hat x}\left.\frac{\partial x_j}{\partial y_J}\right\vert_{\hat x}
,
\end{equation}
where we used eq.~(\ref{equ:dStilde}) and defined the $N\times N$ unconstrained modified Hessian matrix $\tilde{M}_\kappa$ with elements
\begin{equation}
    \tilde{M}_{\kappa,ij}=\left.\frac{\partial^2 \tilde{S}_\kappa}{\partial x_i \partial x_j}\right\vert_{\hat x},
\end{equation}
in analogy with eq.~(\ref{eq:fluctopkappa}).

By definition the constraint function $\xi(\vec{x})$ is constant on the constraint surface {$\mathcal{F}_{\bar\xi}$}, so we have
\begin{equation}\label{eq:constraint first derivative}
    \frac{\partial \xi}{\partial y_I}=\frac{\partial \xi}{\partial x_i}\frac{\partial x_i}{\partial y_I}=0~,
\end{equation}
and
\begin{equation}
    \frac{\partial^2 \xi}{\partial y_I \partial y_J}=\frac{\partial^2 \xi}{\partial x_i \partial x_j}\frac{\partial x_i}{\partial y_I}\frac{\partial x_j}{\partial y_J} + \frac{\partial \xi}{\partial x_i}\frac{\partial^2 x_i}{\partial y_I \partial y_J}=0~,
\end{equation}
everywhere on the constraint surface.

This implies
\begin{equation}
\left.\frac{\partial^2\tilde{S}_\kappa}{\partial y_I \partial y_J}\right\vert_{\hat x}
=
    \left.\frac{\partial(S+\kappa \xi)}{\partial y_I\partial y_J}\right\vert_{\hat x}=\left.\frac{\partial S}{\partial y_I\partial y_J}\right\vert_{\hat x}=M_{\bar\xi,IJ}~.
\end{equation}
Therefore, using eq.~(\ref{eq:secderiv2}), we have
\begin{equation}
    M_{\bar\xi,IJ}=\tilde{M}_{\kappa,ij}
\left.\frac{\partial x_i}{\partial y_I}\right\vert_{\hat x}\left.\frac{\partial x_j}{\partial y_J}\right\vert_{\hat x}~.
\end{equation}
This means that the constrained Hessian $M_{\bar\xi}$ can be obtained by a linear projection of the unconstrained modified Hessian $\tilde{M}_{\kappa}$ to the tangent space of the constraint surface.

Let us now assume that the coordinate system $\{x_i\}$ is oriented in such a way that 
\begin{equation}
    \left.\frac{\partial \xi}{\partial x_i}\right|_{\hat{x}}=0~\text{for}~i\ne N~.
\end{equation}
In a neighbourhood of $\hat{x}$, we can then choose the coordinates $\{y_I\}$ in such a way that
\begin{equation}
    \left.\frac{\partial x_i}{\partial y_I}\right\vert_{\hat x}=\delta_{i,I}~.
\end{equation}
With this choice of coordinates, $M_{\bar\xi}$ is then simply the submatrix of $\tilde{M}_\kappa$ obtained by deleting the $N$th row and the $N$th column, and its determinant is the corresponding cofactor $\tilde{m}_{NN}$ of $\tilde{M}_\kappa$. A standard linear algebra result allows us to express the inverse of $\tilde{M}_\kappa$ in terms of its determinant and cofactors $\tilde{m}_{ij}$ as
\begin{equation}
    \left(\tilde{M}_\kappa^{-1}\right)_{ij}=\frac{\tilde{m}_{ij}}{\det\tilde{M}_\kappa}~,
\end{equation}
and this implies 
\begin{equation}
    {\det}_{\bar\xi} M_{\bar\xi}= \left(\tilde{M}_\kappa^{-1}\right)_{NN}\det\tilde{M}_\kappa~,
\end{equation}
where the subscript in ${\det}_{\bar\xi}$ indicates that this determinant is defined in the $(N-1)$-dimensional tangent space of $\mathcal{F}_{\bar\xi}$~.
Defining the vector $\zeta$ as
\begin{equation}
\label{equ:zetadef0}
    \zeta_i=\left.\frac{\partial \xi}{\partial x_i}\right|_{\hat{x}}~,
\end{equation}
we can express this in a coordinate-independent way as
\begin{equation}
\label{equ:detrelation}
        {\det}_{\bar\xi} M_{\bar\xi}= \nu(\bar\xi) \det\tilde{M}_\kappa~,
\end{equation}
where
\begin{equation}
    \label{equ:nudef}
    \nu(\bar\xi)=\frac{\zeta^T\tilde{M}_\kappa^{-1}\zeta}{\zeta^T\zeta}~.
\end{equation}

To make contact with section~\ref{sec:constinst} we generalise eq.~(\ref{equ:detrelation}) to function spaces, replacing vectors $x_i$ with functions $\phi(x)$ and the saddle point $\hat{x}$ with the solution $\phi_{\bar\xi}(x)$.
There is, however, the complication that our operator $\tilde{M}_\kappa$ has zero eigenvalues associated with translations and therefore it is not invertible. In principle we can project them out and consider only the image space of $\tilde{M}_\kappa$, i.e., the vector space spanned by the eigenfunctions with non-zero eigenvalues. This gives
\begin{equation}
        \label{equ:detrelation2}
        \overline{\text{Det}}_{\bar\xi}\,M_{\bar\xi}=\nu(\bar\xi)\,\overline{\text{Det}}\,\tilde{M}_\kappa~,
\end{equation}
where $\overline{\text{Det}}$ has the same meaning as in eq.~(\ref{equ:saddle2}).
In this image space, $\tilde{M}_\kappa$ is, of course, invertible, and therefore the functional generalisation of eq.~(\ref{equ:nudef}) holds.

For our purposes it is, however, more convenient to instead define a function $\psi$ that satisfies the equation
\begin{equation}
    \label{equ:psidefA}
   \int d^4y \, \tilde{M}_\kappa(x,y)\psi(y)=\zeta(x)~,
\end{equation}
where $\zeta(x)$ is the functional generalisation of eq.~(\ref{equ:zetadef0}),
\begin{equation}
    \zeta(x)=\left.\frac{\delta\xi}{\delta\phi(x)}\right|_{\phi=\phi_{\bar\xi}}
    =\left.\frac{d\mathcal{O}}{d\phi}
    \right|_{\phi=\phi_{\bar\xi}(x)}~.
\end{equation}

When $\tilde{M}_\kappa$ has zero eigenvalues $\psi$ is not unique, but the difference $\Delta\psi(x)$ between any two solutions is a translation and may be written as
\begin{equation}
    \Delta\psi(x)=a_\mu\frac{\partial\phi_{\bar\xi}}{\partial x_\mu},
\end{equation}
where $a_\mu$ are real constants. This means
\begin{equation}
\zeta^T\Delta\psi=\int d^4x \zeta(x)\Delta\psi(x)
=a_\mu \int d^4x \left.\frac{d\mathcal{O}}{d\phi}
    \right|_{\phi=\phi_{\bar\xi}(x)} \frac{\partial\phi_{\bar\xi}}{\partial x_\mu}
=a_\mu\int d^4x \frac{\partial\mathcal{O}(\phi_{\bar\xi}(x))}{\partial x_\mu}=0~.
\end{equation}
Therefore we can write eq.~(\ref{equ:nudef}) in a functional form as
\begin{equation}
\label{equ:nufunc}
    \nu(\bar\xi)=\frac{\int d^4x \zeta(x)\psi(x)}{\int d^4x \zeta(x)^2}~,
\end{equation}
where $\psi$ is any function that satisfies eq.~(\ref{equ:psidefA}).

\section{Constrained instanton solutions --- consistency checks}\label{app:CISols} 

There exist multiple ways of checking whether the field configurations we found are indeed the constrained instantons we were looking for. These checks make use of the properties of Lagrange multipliers described in section~\ref{sec:Lagmult}. We performed three separate checks on our solutions. 

\begin{figure}[t]
    \centering
    \includegraphics[width=0.45\linewidth]{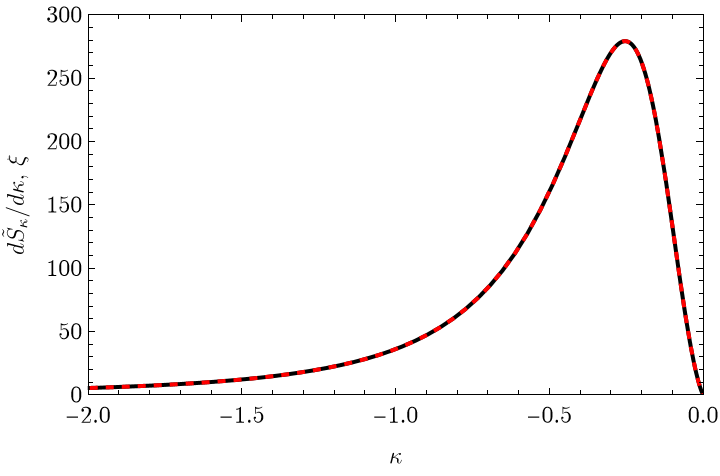}
    \qquad
    \includegraphics[width=0.475\textwidth]{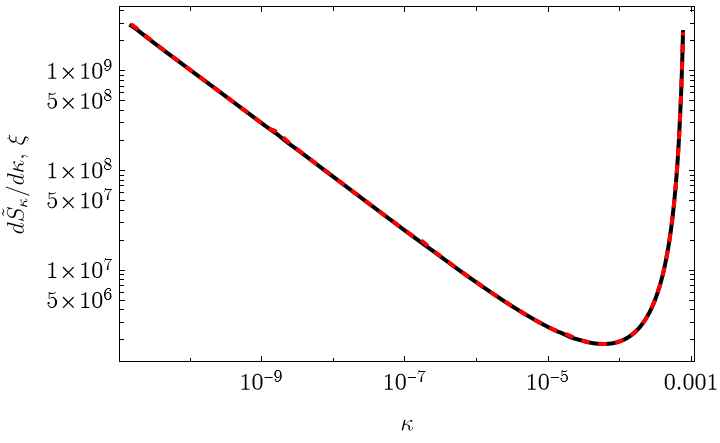}
    \caption{A comparison between the absolute value of the derivative of the unconstrained action with respect to the Lagrange multiplier (black) and the calculated values of the constraint (red dashed). Left: $\phi^3$ constraint. Right: $\phi^6$ constraint.}
    \label{fig:legendrechecks}
\end{figure}

The first check is based on eq.~(\ref{eq:legendrexi}), reproduced here for convenience:
\begin{equation}
    \bar\xi=\frac{d\tilde S_\kappa(\kappa)}{d\kappa}~.
\end{equation}
We numerically computed $d\tilde S_\kappa(\kappa)/d\kappa$ and compared it with the numerically obtained values of the constraint and found excellent agreement. This is shown for both types of constraint in figure~\ref{fig:legendrechecks}.

A different kind of check is based on the fact that constrained instantons are stationary points of the modified action (eq.~\eqref{eq:deformact}) \emph{with respect to scaling transformations}. By considering the behaviour of the modified action under two different scaling transformations, we obtained two different integral identities that a constrained instanton solution must satisfy.

\begin{figure}[t]
    \centering
    \includegraphics[width=0.45\textwidth]{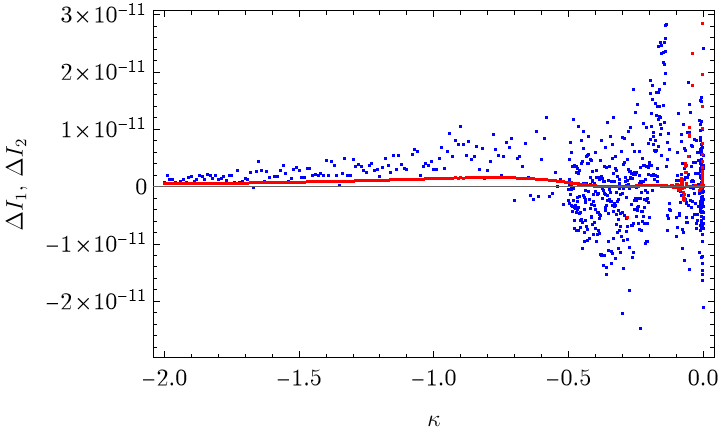}
    \qquad
    \includegraphics[width=0.45\textwidth]{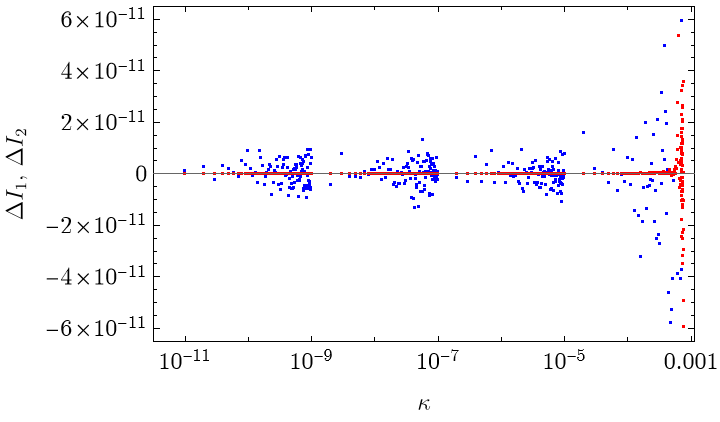}
    \caption{$\Delta I_1$ (blue) and $\Delta I_2$ (red) of eq.~\eqref{eq:integralchecks} as a function of $\kappa$. Left: $\phi^3$ constraint. Right: $\phi^6$ constraint.}
    \label{fig:integralchecks}
\end{figure}

First, consider the following scaling transformation
\begin{equation}
    \phi(x)\rightarrow \phi(ax)~.
\end{equation}
Under this transformation the modified action in eq.~\eqref{eq:deformact} becomes
\begin{equation}\label{eq:scacttriv}
    \tilde S^{(a)}_\kappa=\int d^4x\,\left(a^{-2}\frac{1}{2}(\partial_\mu\phi)^2+a^{-4}\frac{1}{2}m^2\phi^2-a^{-4}\frac{\lambda}{4!}\phi^4\right)+a^{-4}\kappa\xi[\phi]~.
\end{equation}
Demanding the constrained instanton extremises the unconstrained action at $a=1$ we get
\begin{equation}
    \kappa\xi[\phi]=-\int d^4x \left(\frac{1}{4}(\partial_\mu\phi)^2+\frac{1}{2}m^2\phi^2-\frac{\lambda}{4!}\phi^4\right) = I_1[\phi]~.
\end{equation}
Next, consider a different scaling transformation
\begin{equation}
    \phi(x)\rightarrow a\phi(ax)~.
\end{equation}
Under the above transformation, the modified action of eq.~\eqref{eq:deformact} becomes
\begin{equation}
    \tilde S_\kappa^{(a)}= \int d^4x \left(\frac{1}{2}(\partial_\mu\phi)^2+a^{-2}\frac{1}{2}m^2\phi^2-\frac{\lambda}{4!}\phi^4\right)+a^{d-4}\kappa\xi[\phi]~.
\end{equation}
Demanding that the action is stationary with respect to $a$ at $a=1$ we obtain
\begin{equation}
    \kappa\xi[\phi]=\frac{1}{d-4}\int d^4x\,m^2\phi^2=I_2[\phi]~.
\end{equation}
We computed the integrals $I_1$ and $I_2$ numerically and compared them with the calculated values of $\kappa\bar\xi$. We define the following quantities
\begin{equation}\label{eq:integralchecks}
    \begin{array}{c}
         \Delta I_1=I_1[\phi]-\kappa\xi[\phi]~, \\
             \Delta I_2=I_2[\phi]-\kappa\xi[\phi]~.
    \end{array}
\end{equation}
These quantities for both constraints (for $m=\lambda=1$) are plotted in figure~\ref{fig:integralchecks}.

\section{Numerics --- consistency checks}\label{app:numcheck}

We carried out several checks to ensure the numerics were robust and our results were independent of the size of the simulation interval and chosen working precision. 
Our default simulation box size for most of the solutions for both constraints was determined by the minimum radius of $R_{\min}=10^{-7}$ and the maximum radius of $R_{\max}=50$, with some exceptions as explained in sections \ref{sec:phi3} and \ref{sec:phi6}. In this appendix, we set $m=\lambda=1$, without loss of generality, as explained in section~\ref{sec:params}.

First, we tested the robustness of our calculation with respect to the variation of the simulation box size. We picked several values of $\kappa$ for each constraint and repeated the calculation of the constrained action --- starting from finding the constrained instanton solution --- for different values of $R_{\max}$, while keeping $R_{\min}$ constant at the default value. We repeated the same calculation for different values of $R_{\min}$, while keeping $R_{\max}$ fixed at the default value. For the $\phi^3$ constraint, we picked 2 values of $\kappa$, $K_3 = -10^{-2}$ (upper branch of $S(\xi)$) and $K_3 = -1$ (lower branch of $S(\xi)$). For the $\phi^6$ constraint, we chose three values of $\kappa$, $K_6 = 10^{-6}$ (upper branch of $S(\xi)$), $K_6 =10^{-4}$ (lower branch of $S(\xi)$, $S>0$), and $K_6 = 5\times10^{-4}$ (lower branch of $S(\xi)$, $S<0$).

\begin{figure}[t]
    \centering
    \includegraphics[width=0.4\textwidth]{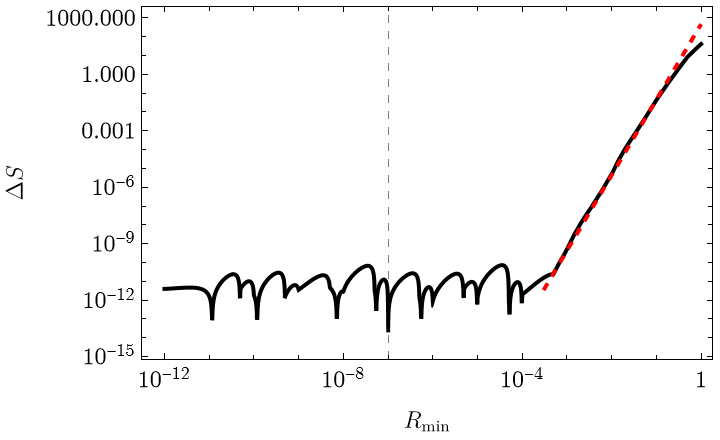}
    \qquad
    \includegraphics[width=0.4\textwidth]{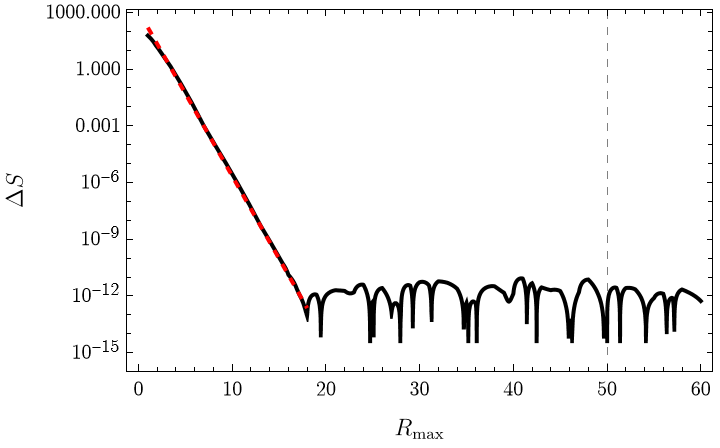}
    \caption{$\phi^3$ constraint. The difference between the value of the action computed at $R_{\min}=10^{-7}$ and $R_{\max}=50$ for $K_3 = -1$, and the action computed for different values of $R_{\min}$ (left) and different values of $R_{\max}$ (right).}
    \label{fig:phi3rmaxrmin}
\end{figure}

\begin{figure}[t]
    \centering
    \includegraphics[width=0.4\textwidth]{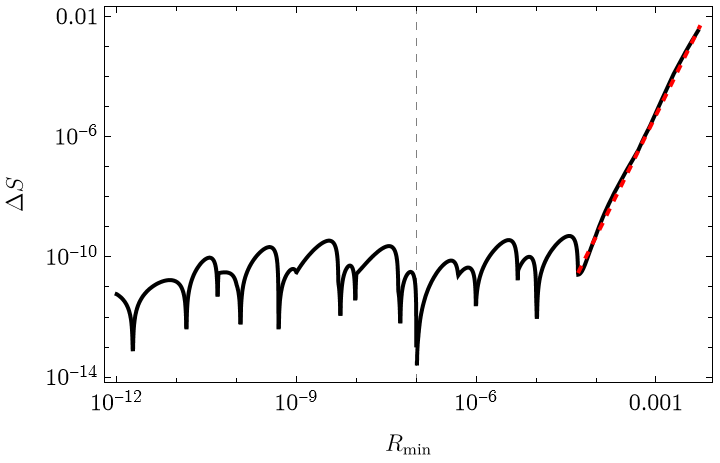}
    \qquad
    \includegraphics[width=0.4\textwidth]{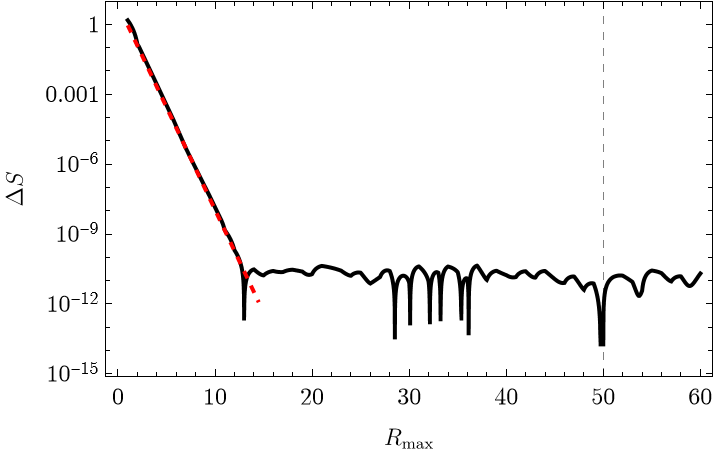}
    \caption{$\phi^6$ constraint. The difference between the value of the action computed at $R_{\min}=10^{-7}$ and $R_{\max}=50$ for $K_6 = 10^{-6}$, and the action computed for different values of $R_{\min}$ (left) and $R_{\max}$ (right).}
    \label{fig:phi6rmaxrmin}
\end{figure}

The effects these variations on the action are shown in figure~\ref{fig:phi3rmaxrmin} ($\phi^3$ constraint) and figure~\ref{fig:phi6rmaxrmin} ($\phi^6$ constraint). In all cases we see that our chosen default values of the simulation box boundaries (indicated by the vertical dashed lines) are well within the region of the parameter space dominated by random numerical errors, and therefore our calculation is robust with respect to the variation of the simulation box size. In both cases we were able to determine the functional form of $\Delta S$ as it approaches the random-error-dominated region
\begin{align} \nonumber
    \Delta S_{R_{\max}} &\sim a e^{-b R_{\max}}~, \\
    \Delta S_{R_{\min}} &\sim c \,(R_{\min})^d~.
    \label{error-fit-params}
\end{align}

We were also able to determine the exponents and prefactors, $a$, $b$, $c$, and $d$. These are shown in tables~\ref{tab:phi3errors} and~\ref{tab:phi6errors} . It seems that for both constraints and all values of $\kappa$, the value of $b$ is consistent with $2$, and the value of $d$ is consistent with $4$. The prefactors $a$ and $c$ are all different, which is not surprising --- it is expected that the prefactors will be heavily dependent on the corresponding instanton size, and therefore on the value of $\kappa$.

\begin{table}[t]
    \centering
     \begin{tabular}{|c|c|c|c|c|}
        \hline
        $\kappa$ & $a$ & $b$ & $c$ & $d$ \\
        \hline
        $-10^{-2}$ & $(4.5 \pm 0.2)\times10^{-2}$ & $2.04 \pm 0.04$ & $(9.9 \pm 0.4)\times10^{12}$ & $3.97 \pm 0.04$ \\
        $-1$ & $(11.8 \pm 0.2)\times 10^2$ & $2.01 \pm 0.02$ & $(4.7 \pm 0.3) \times 10^2$ & $4.01 \pm 0.07$ \\
        \hline
    \end{tabular}
    \caption{Error fit parameters for $R_{\max}$ and $R_{\min}$ (eq.~\eqref{error-fit-params}), for the $\phi^3$ constraint.}
    \label{tab:phi3errors}
\end{table}

We also checked that our results are stable with respect to changing the numerical working precision used to carry out the calculations. This was carried out analogously to the procedure described above, repeating the calculation of the constrained action for different values of $\kappa$ for different  working precision.

\begin{table}[t]
    \centering
     \begin{tabular}{|c|c|c|c|c|}
        \hline
        $\kappa$ & $a$ & $b$ & $c$ & $d$ \\
        \hline
        $10^{-6}$ & $7.0 \pm 0.4$ & $2.03 \pm 0.06$ & $(6.7 \pm 0.7)\times10^{6}$ & $4.0 \pm 0.1$ \\
        $10^{-4}$ & $(2.1 \pm 0.1)\times10^{3}$ & $2.02 \pm 0.05$ & $(3.09 \pm 0.05)\times10^{3}$ & $4.01 \pm 0.02$ \\
        $5\times10^{-4}$ & $(1.00 \pm 0.05)\times10^{8}$ & $2.01 \pm 0.05$ & $2.8 \pm 0.4$ & $4.1 \pm 0.2$ \\
        \hline
    \end{tabular}
    \caption{Error fit parameters for $R_{\max}$ and $R_{\min}$, for the $\phi^3$ constraint.}
    \label{tab:phi6errors}
\end{table}
\begin{figure}[t]
    \centering
    \includegraphics[width=0.4\textwidth]{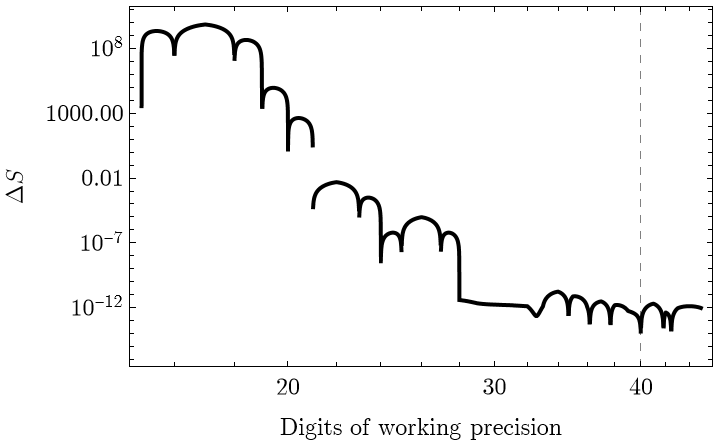}
    \qquad
    \includegraphics[width=0.4\textwidth]{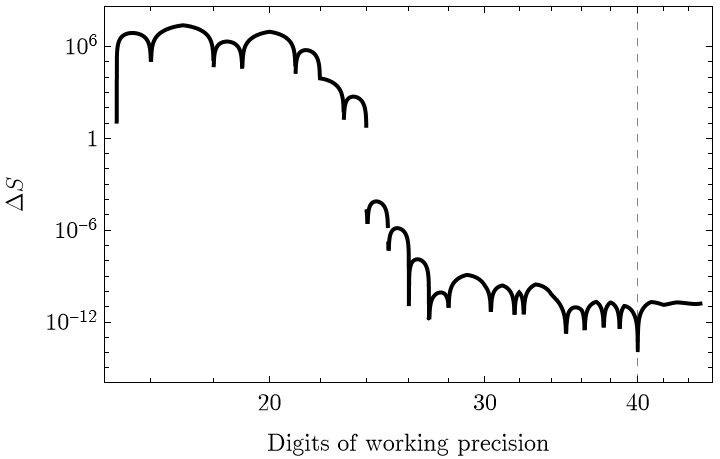}
    \caption{The difference between the action computed at a working precision of 40 digits and the action computed for different values of the working precision. Left: $\phi^3$ constraint, $K_3 = -1$. Right: $\phi^6$ constraint, $K_6 = 10^{-6}$.}
    \label{fig:phiwperr}
\end{figure}

As can be seen in figure~\ref{fig:phiwperr}, the difference between the computed action and the reference action is much more irregular than when varying simulation box size. However, there is still a clearly discernable overall trend, and we can see that the error improves as working precision increases. It is also clear that our choice for the working precision is again well within the regime dominated by random numerical error, rather than any systematic error.

 
\providecommand{\href}[2]{#2}\begingroup\raggedright\endgroup


\begin{thebibliography}{10}

\bibitem{Coleman:1977py}
S.R.~Coleman, \emph{{The Fate of the False Vacuum. 1. Semiclassical Theory}}, \href{https://doi.org/10.1103/PhysRevD.16.1248}{\emph{Phys. Rev. D} {\bfseries 15} (1977) 2929}.

\bibitem{Callan:1977pt}
C.G.~Callan, Jr. and S.R.~Coleman, \emph{{The Fate of the False Vacuum. 2. First Quantum Corrections}}, \href{https://doi.org/10.1103/PhysRevD.16.1762}{\emph{Phys. Rev. D} {\bfseries 16} (1977) 1762}.

\bibitem{Affleck:1980mp}
I.~Affleck, \emph{{On Constrained Instantons}}, \href{https://doi.org/10.1016/0550-3213(81)90307-2}{\emph{Nucl. Phys. B} {\bfseries 191} (1981) 429}.

\bibitem{Belavin:1975fg}
A.A.~Belavin, A.M.~Polyakov, A.S.~Schwartz and Y.S.~Tyupkin, \emph{{Pseudoparticle Solutions of the Yang-Mills Equations}}, \href{https://doi.org/10.1016/0370-2693(75)90163-X}{\emph{Phys. Lett. B} {\bfseries 59} (1975) 85}.

\bibitem{Adler:1969gk}
S.L.~Adler, \emph{{Axial vector vertex in spinor electrodynamics}}, \href{https://doi.org/10.1103/PhysRev.177.2426}{\emph{Phys. Rev.} {\bfseries 177} (1969) 2426}.

\bibitem{Bell:1969ts}
J.S.~Bell and R.~Jackiw, \emph{{A PCAC puzzle: $\pi^0 \to \gamma \gamma$ in the $\sigma$ model}}, \href{https://doi.org/10.1007/BF02823296}{\emph{Nuovo Cim. A} {\bfseries 60} (1969) 47}.

\bibitem{tHooft:1976rip}
G.~'t~Hooft, \emph{{Symmetry Breaking Through Bell-Jackiw Anomalies}}, \href{https://doi.org/10.1103/PhysRevLett.37.8}{\emph{Phys. Rev. Lett.} {\bfseries 37} (1976) 8}.

\bibitem{Derrick:1964ww}
G.H.~Derrick, \emph{{Comments on nonlinear wave equations as models for elementary particles}}, \href{https://doi.org/10.1063/1.1704233}{\emph{J. Math. Phys.} {\bfseries 5} (1964) 1252}.

\bibitem{Fubini:1976jm}
S.~Fubini, \emph{{A New Approach to Conformal Invariant Field Theories}}, \href{https://doi.org/10.1007/BF02785664}{\emph{Nuovo Cim. A} {\bfseries 34} (1976) 521}.

\bibitem{Lee:1985uv}
K.-M.~Lee and E.J.~Weinberg, \emph{{TUNNELING WITHOUT BARRIERS}}, \href{https://doi.org/10.1016/0550-3213(86)90150-1}{\emph{Nucl. Phys. B} {\bfseries 267} (1986) 181}.

\bibitem{tHooft:1976snw}
G.~'t~Hooft, \emph{{Computation of the Quantum Effects Due to a Four-Dimensional Pseudoparticle}}, \href{https://doi.org/10.1103/PhysRevD.14.3432}{\emph{Phys. Rev. D} {\bfseries 14} (1976) 3432}.

\bibitem{PhysRevD.19.540}
Y.~Frishman and S.~Yankielowicz, \emph{Large-order behavior of perturbation theory and mass terms}, \href{https://doi.org/10.1103/PhysRevD.19.540}{\emph{Phys. Rev. D} {\bfseries 19} (1979) 540}.

\bibitem{Silvestrov:1992ct}
P.G.~Silvestrov, \emph{{Constrained instanton and baryon number nonconservation at high-energies}}, \href{https://doi.org/10.1016/0370-2693(94)00059-X}{\emph{Phys. Lett. B} {\bfseries 323} (1994) 25} [\href{https://arxiv.org/abs/hep-ph/9212215}{{\ttfamily hep-ph/9212215}}].

\bibitem{Espinosa:1989qn}
O.~Espinosa, \emph{{High-Energy Behavior of Baryon and Lepton Number Violating Scattering Amplitudes and Breakdown of Unitarity in the Standard Model}}, \href{https://doi.org/10.1016/0550-3213(90)90473-Q}{\emph{Nucl. Phys. B} {\bfseries 343} (1990) 310}.

\bibitem{osti_6099255}
P.B.~Arnold and M.P.~Mattis, \emph{Baryon-number violation with improved instantons}, \href{https://doi.org/10.1103/PhysRevLett.66.13}{\emph{Physical Review Letters; (USA)} {\bfseries 66:1} (1991) }.

\bibitem{Ringwald:1989ee}
A.~Ringwald, \emph{{High-Energy Breakdown of Perturbation Theory in the Electroweak Instanton Sector}}, \href{https://doi.org/10.1016/0550-3213(90)90300-3}{\emph{Nucl. Phys. B} {\bfseries 330} (1990) 1}.

\bibitem{Khoze:1990bm}
V.V.~Khoze and A.~Ringwald, \emph{{Total cross-section for anomalous fermion number violation via dispersion relation}}, \href{https://doi.org/10.1016/0550-3213(91)90118-H}{\emph{Nucl. Phys. B} {\bfseries 355} (1991) 351}.

\bibitem{Klinkhamer:1991pq}
F.R.~Klinkhamer, \emph{{On the existence of a new constrained instanton and high-energy electroweak interactions}}, \href{https://doi.org/10.1016/0550-3213(92)90125-U}{\emph{Nucl. Phys. B} {\bfseries 376} (1992) 255}.

\bibitem{Klinkhamer:1993kn}
F.R.~Klinkhamer, \emph{{Existence of a new instanton in constrained Yang-Mills Higgs theory}}, \href{https://doi.org/10.1016/0550-3213(93)90275-T}{\emph{Nucl. Phys. B} {\bfseries 407} (1993) 88} [\href{https://arxiv.org/abs/hep-ph/9306208}{{\ttfamily hep-ph/9306208}}].

\bibitem{Klinkhamer:1994xw}
F.R.~Klinkhamer, \emph{{New electroweak instanton and possible breakdown of unitarity}},  in \emph{{27th International Conference on High-energy Physics}}, 7, 1994 [\href{https://arxiv.org/abs/hep-ph/9409336}{{\ttfamily hep-ph/9409336}}].

\bibitem{Klinkhamer:1996ad}
F.R.~Klinkhamer and J.~Weller, \emph{{Construction of a new constrained instanton in Yang-Mills Higgs theory}}, \href{https://doi.org/10.1016/S0550-3213(96)90145-5}{\emph{Nucl. Phys. B} {\bfseries 481} (1996) 403} [\href{https://arxiv.org/abs/hep-ph/9606481}{{\ttfamily hep-ph/9606481}}].

\bibitem{Klinkhamer:1997pd}
F.R.~Klinkhamer, \emph{{Fermion zero modes of a new constrained instanton in Yang-Mills Higgs theory}}, \href{https://doi.org/10.1016/S0550-3213(97)00828-6}{\emph{Nucl. Phys. B} {\bfseries 517} (1998) 142} [\href{https://arxiv.org/abs/hep-th/9709194}{{\ttfamily hep-th/9709194}}].

\bibitem{Gibbs:1995xt}
M.J.~Gibbs, \emph{{Electroweak instantons at nonzero Weinberg angle}}, \href{https://doi.org/10.1016/0370-2693(95)00126-6}{\emph{Phys. Lett. B} {\bfseries 348} (1995) 149} [\href{https://arxiv.org/abs/hep-ph/9501253}{{\ttfamily hep-ph/9501253}}].

\bibitem{PhysRevD.61.105020}
M.~Nielsen and N.K.~Nielsen, \emph{Explicit construction of constrained instantons}, \href{https://doi.org/10.1103/PhysRevD.61.105020}{\emph{Phys. Rev. D} {\bfseries 61} (2000) 105020}.

\bibitem{Aoyama:1995ca}
H.~Aoyama, T.~Harano, M.~Sato and S.~Wada, \emph{{Valley instanton versus constrained instanton}}, \href{https://doi.org/10.1016/0550-3213(96)00066-1}{\emph{Nucl. Phys. B} {\bfseries 466} (1996) 127} [\href{https://arxiv.org/abs/hep-th/9512064}{{\ttfamily hep-th/9512064}}].

\bibitem{Aoyama:1995zt}
H.~Aoyama, T.~Harano, M.~Sato and S.~Wada, \emph{{Valley instanton in the gauge Higgs system}}, \href{https://doi.org/10.1142/S0217732396000072}{\emph{Mod. Phys. Lett. A} {\bfseries 11} (1996) 43} [\href{https://arxiv.org/abs/hep-th/9507111}{{\ttfamily hep-th/9507111}}].

\bibitem{Wang:1994rz}
M.-y.~Wang, \emph{{On constrained instanton valleys}},  \href{https://arxiv.org/abs/hep-ph/9502333}{{\ttfamily hep-ph/9502333}}.

\bibitem{Espinosa:2018hue}
J.R.~Espinosa, \emph{{A Fresh Look at the Calculation of Tunneling Actions}}, \href{https://doi.org/10.1088/1475-7516/2018/07/036}{\emph{JCAP} {\bfseries 07} (2018) 036} [\href{https://arxiv.org/abs/1805.03680}{{\ttfamily 1805.03680}}].

\bibitem{Espinosa:2019hbm}
J.R.~Espinosa, \emph{{Tunneling without Bounce}}, \href{https://doi.org/10.1103/PhysRevD.100.105002}{\emph{Phys. Rev. D} {\bfseries 100} (2019) 105002} [\href{https://arxiv.org/abs/1908.01730}{{\ttfamily 1908.01730}}].

\bibitem{Espinosa:2022jlx}
J.R.~Espinosa and J.F.~Fortin, \emph{{Vacuum decay actions from tunneling potentials for general spacetime dimension}}, \href{https://doi.org/10.1088/1475-7516/2023/02/023}{\emph{JCAP} {\bfseries 02} (2023) 023} [\href{https://arxiv.org/abs/2211.13667}{{\ttfamily 2211.13667}}].

\bibitem{Espinosa:2024ufg}
J.R.~Espinosa, \emph{{The Unreasonable Effectiveness of the Tunneling Potential}}, \href{https://doi.org/10.22323/1.463.0164}{\emph{PoS} {\bfseries CORFU2023} (2024) 164} [\href{https://arxiv.org/abs/2404.19657}{{\ttfamily 2404.19657}}].

\bibitem{Espinosa:2025ejf}
J.R.~Espinosa and T.~Konstandin, \emph{{An Exploration of Vacuum-Decay Valleys}},  \href{https://arxiv.org/abs/2506.06154}{{\ttfamily 2506.06154}}.

\bibitem{PhysRevD.104.L081501}
J.~Cotler and K.~Jensen, \emph{Gravitational constrained instantons}, \href{https://doi.org/10.1103/PhysRevD.104.L081501}{\emph{Phys. Rev. D} {\bfseries 104} (2021) L081501}.

\bibitem{GarciaMartin-Caro:2022ukd}
A.~Garc{\'\i}a Mart{\'\i}n-Caro, \emph{{Constrained instanton approximation of Skyrmions with massive pions}}, \href{https://doi.org/10.1016/j.physletb.2022.137532}{\emph{Phys. Lett. B} {\bfseries 835} (2022) 137532} [\href{https://arxiv.org/abs/2209.06607}{{\ttfamily 2209.06607}}].

\bibitem{Press:2007ipz}
W.H.~Press, S.A.~Teukolsky, W.T.~Vetterling and B.P.~Flannery, \emph{{Numerical Recipes: The Art of Scientific Computing (Third Edition)}}, Cambridge University Press (2007).

\bibitem{doi:10.1137/1.9781611971170}
R.~Bellman, \emph{Introduction to Matrix Analysis, Second Edition}, Society for Industrial and Applied Mathematics, second~ed. (1997), \href{https://doi.org/10.1137/1.9781611971170}{10.1137/1.9781611971170}, [\href{https://arxiv.org/abs/https://epubs.siam.org/doi/pdf/10.1137/1.9781611971170}{{\ttfamily https://epubs.siam.org/doi/pdf/10.1137/1.9781611971170}}].

\bibitem{asymptotic-limits}
B.~Elder, K.~Gawrych and A.~Rajantie, ``Asymptotic limits of constrained instantons.'' In preparation.

\bibitem{Degrassi:2012ry}
G.~Degrassi, S.~Di~Vita, J.~Elias-Miro, J.R.~Espinosa, G.F.~Giudice, G.~Isidori et~al., \emph{{Higgs mass and vacuum stability in the Standard Model at NNLO}}, \href{https://doi.org/10.1007/JHEP08(2012)098}{\emph{JHEP} {\bfseries 08} (2012) 098} [\href{https://arxiv.org/abs/1205.6497}{{\ttfamily 1205.6497}}].

\bibitem{Markkanen:2018pdo}
T.~Markkanen, A.~Rajantie and S.~Stopyra, \emph{{Cosmological Aspects of Higgs Vacuum Metastability}}, \href{https://doi.org/10.3389/fspas.2018.00040}{\emph{Front. Astron. Space Sci.} {\bfseries 5} (2018) 40} [\href{https://arxiv.org/abs/1809.06923}{{\ttfamily 1809.06923}}].

\bibitem{Branchina:2014rva}
V.~Branchina, E.~Messina and M.~Sher, \emph{{Lifetime of the electroweak vacuum and sensitivity to Planck scale physics}}, \href{https://doi.org/10.1103/PhysRevD.91.013003}{\emph{Phys. Rev. D} {\bfseries 91} (2015) 013003} [\href{https://arxiv.org/abs/1408.5302}{{\ttfamily 1408.5302}}].

\end{thebibliography}
\end{document}